\documentclass[prx,twocolumn,nofootinbib]{revtex4-1}

\usepackage{amsmath,amssymb,amsfonts,amsthm}
\usepackage{amsmath}
\usepackage{tikz-cd}
\usepackage{quiver}
\usepackage{float}
\usepackage{dcolumn}
\usepackage{epsfig}
\usepackage{graphicx}
\usepackage{graphics}
\usepackage[latin1]{inputenc}
\usepackage{latexsym}
\usepackage{rotating}
\usepackage{xcolor}
\usepackage{slashed}
\usepackage{setspace}
\usepackage{mathtools}
\definecolor{Burgundy}{RGB}{144,0,32}
\definecolor{Burgundy1}{RGB}{128,0,32}
\definecolor{Burgundy2}{RGB}{158,5,8}
\definecolor{VividBurgundy}{RGB}{159,29,53}
\usepackage{braket}

\allowdisplaybreaks

\usepackage{hyperref}
\hypersetup{
    colorlinks=true,
    citecolor=VividBurgundy,
    filecolor=black,
    linkcolor=VividBurgundy,
    urlcolor=VividBurgundy
}

\newcolumntype{L}{>{$}c<{$}} 

\renewcommand{\bm}{\textbf}

\newcommand{\Z}{\mathbb{Z}}
\newcommand{\R}{\mathbb{R}}
\newcommand{\C}{\mathbb{C}}
\newcommand{\cC}{\mathcal{C}}

\newcommand{\cH}{\mathcal{H}}
\newcommand{\cM}{\mathcal{M}}
\newcommand{\cT}{\mathcal{T}}
\newcommand{\cZ}{\mathcal{Z}}
\DeclareMathOperator{\tr}{tr}
\DeclareMathOperator{\Str}{\textbf{Str}}
\DeclareMathOperator{\Rep}{\textbf{Rep}}
\DeclareMathOperator{\Hom}{Hom}
\DeclareMathOperator{\QCD}{QCD_2}

\begin{document}

\title{Particle-Soliton Degeneracy in 2D Quantum Chromodynamics}

\author{Clay C\'{o}rdova} \email{clayc@uchicago.edu}

\author{Diego Garc\'{i}a-Sep\'{u}lveda} \email{dgarciasepulveda@uchicago.edu}

\author{Nicholas Holfester} \email{nholfester@uchicago.edu}

\affiliation{Kadanoff Center for Theoretical Physics  \& Enrico Fermi Institute, University of Chicago}

\begin{abstract}
\noindent 
Quantum chromodynamics in two spacetime dimensions admits a finite non-invertible symmetry described mathematically by a fusion category. This symmetry is spontaneously broken at long distances, leading to distinct vacua. When the theory has a mass gap, the spectrum is therefore characterized by particle excitations above a single vacuum and soliton sectors interpolating between vacua. We use anyon condensation and the representation theory of fusion categories to obtain exact results about this spectrum, exhibiting the allowed multiplets. Often, particles and solitons are in the same representation and therefore must have equal masses. Furthermore, the fusion category symmetry frequently implies the existence of certain stable states in the spectrum. The resulting degeneracies are encoded in quiver diagrams where nodes are vacua and arrows are excited states. 
\end{abstract}

\maketitle

\begin{spacing}{.5}
\tableofcontents
\end{spacing}

\section{Introduction}

Gauge theory is a central paradigm in particle physics.  Among its most fascinating properties is its strongly coupled nature: ultraviolet (UV) degrees of freedom consisting of manifest quarks and gauge fields give way in the infrared (IR) to a surprising particle spectrum of mesons and baryons. Analytically understanding this phenomenon has served as a driving goal of theoretical physics for the past half-century. 

Two-dimensional (2D) theories have long served as a playground for testing new techniques to attack strongly-coupled physics.  In particular, in quantum chromodynamics (QCD$_{2}$) 't Hooft famously solved the theory in the large number of colors limit \cite{tHooft:1974pnl}. More recently, novel non-invertible symmetries have been applied to understand confinement \cite{Komargodski:2020mxz}, and to characterize the vacuum structure of QCD$_{2}$ where the distinct states are parameterized by expectation values of gauge-invariant composite operators \cite{Cordova:2024goh, Delmastro:2021otj, Delmastro:2022prj}. (See Appendix \ref{VacuumCondensates} below for explicit examples in $\QCD$.)

In this work, we continue this line of development building on the recent results \cite{ Cordova:2024vsq, Cordova:2024iti}. Focusing on the case of gapped $\QCD$, we aim to directly characterize the spectrum. We realize $\QCD$ via an interval compactification from a three-dimensional (3D) topological quantum field theory (TQFT) and following \cite{Cordova:2024goh} identify the fusion category $\cC$ which describes the mathematical structure of the non-invertible symmetry \cite{Frohlich:2006ch, Frohlich:2009gb, Bhardwaj:2017xup, Chang:2018iay}. Physically, such symmetries are realized by topological line operators  that commute with the Hamiltonian \cite{Gaiotto:2014kfa}, but are not in general represented by unitary operators acting on the Hilbert space. Moreover, as noted in \cite{Cordova:2024goh}, the fusion category $\cC$ is fully spontaneously broken in the IR and hence characterizes the vacua of the gauge theory. 

We harness the representation theory of this fusion category to determine the allowed multiplets of particles, excitations above a single vacua, and solitons interpolating between distinct vacua.  Strikingly, we find that the fusion category symmetry $\cC$ often implies that particle and solitons are in the same representation and hence necessarily have equal masses.  This feature is unique to spontaneously broken non-invertible symmetries and illustrates the intrinsically quantum (strongly-coupled) nature of these symmetries.  

We exhibit our results in several explicit gauge theories below.  The implied degeneracies are elegantly encoded in a quiver diagram where the nodes are vacua and the arrows are excited particle and soliton states. Such diagrams echo those discussed in many contexts such as the lattice integrable models of \cite{DiFrancesco:1989ha}, the relation between graphs, non-diagonal modular invariants, and conformal boundary conditions of \cite{Petkova:1995fw, Behrend:1999bn} (see also \cite{kirillov2002q, Hung:2015hfa} for the relationship between anyon condensation and generalized ADE diagrams), and in the direct analysis of supersymmetric solitons \cite{Cecotti:1992rm}.

\section{Background on QCD$_{2}$}\label{sec:background}

We are interested in QCD in two spacetime dimensions with massless fermions. The gauge group is denoted by $G$, and the fermions transform in a representation $\mathbf{R}$ of $G$.  We take the fermions to have vanishing bare mass. The action is:
\begin{equation} \label{fermionic2DQCD}
    S_{f}(g_{\mathrm{YM}}) = \int_{\Sigma} d^{2}x  \Big[ \mathrm{Tr}(\Psi^{T} i\slashed{D}_{\rho} \Psi)   - \frac{1}{4g_{\mathrm{YM}}^{2}}\ \mathrm{Tr}(F^{2}) \Big].
\end{equation}
For simplicity in the following, we consider only vector-like theories where the left-moving and right-moving fermions transform in the same representation $\mathbf{R}$ of the gauge group $G$, and further restrict to the case where $\mathbf{R}$ is irreducible.

As defined above, $\QCD$ is a fermionic theory, i.e.\ there are fermionic local operators and the partition functions depend on a choice of spacetime spin structure.
To simplify our discussion, we can bosonize the fermions \cite{Witten:1983ar}, and recast these degrees of freedom as a theory of currents, a WZW model $\mathrm{Spin}(\mathrm{dim}(\mathbf{R}))_{1}.$ Coupling to $G$ gauge fields then results in a bosonic version of $\QCD$ as a gauged WZW model with a $G$ gauge field kinetic term:
\begin{equation} \label{bosonic2DQCD}
    S_{b}(g_{\mathrm{YM}}) = S_{\mathrm{WZW}}[g,A] - \frac{1}{4g_{\mathrm{YM}}^{2}} \int_{\Sigma} d^{2}x \ \mathrm{Tr}(F^{2}).
\end{equation}
We present our results below for this bosonized theory. Since bosonization/fermionization 
is an invertible operation \cite{Ji:2019ugf, Thorngren:2021yso}, this involves no loss in generality.

Our goal is to constrain particle and soliton states. This is cleanest when the theory in question has a mass gap so that we can separate the spectrum from any residual gapless modes.  In $\QCD$, there is a simple criterion that controls the gap \cite{Delmastro:2021otj}.  Denote by $I(\mathbf{R})$ the Dynkin index of the representation $\mathbf{R}$ and by $c_{G_{k}}$ the central charge of the WZW model based on a Lie group $G$ at level $k$.  Then, the theory is gapped if and only if the coset
\begin{equation} \label{2DQCD-TopologicalCoset}
    \frac{\mathrm{Spin}( \mathrm{dim}(\mathbf{R}))_{1}}{G_{I(\mathbf{R})}},
\end{equation}
 has vanishing central charge:
\begin{equation} \label{coset-centralcharge}
    c_{\mathrm{Spin}( \mathrm{dim}(\mathbf{R}))_{1}/G_{I(\mathbf{R})}} = c_{\mathrm{Spin}( \mathrm{dim}(\mathbf{R}))_{1}} - c_{G_{I(\mathbf{R})}} = 0.
\end{equation}
 When this is the case, we refer to \eqref{2DQCD-TopologicalCoset} as a \textit{topological coset}. This condition for a gap is valid irrespective of the global form of the gauge group $G$, provided that $\mathbf{R}$ is an allowed representation of $G$.

Next we turn to the effective infrared (IR) description of the $\QCD$ theory \eqref{bosonic2DQCD}. A natural candidate arises if we examine \eqref{bosonic2DQCD}, and assume that the IR is described by the corresponding $g_{\mathrm{YM}} \to \infty$ limit:
\begin{align} \label{gaugedWZW}
    S_{b}(g_{\mathrm{YM}}) \xrightarrow[g_{\mathrm{YM}} \to \infty]{} S_{\mathrm{WZW}}[g,A]. 
\end{align}
We find that the IR is described by the aforementioned gauged WZW model consisting of $\mathrm{Spin}( \mathrm{dim}(\mathbf{R}))_{1}$ matter content coupled to $G$ gauge fields. We assume this description below. Algebraically, this corresponds to the quotient of chiral algebras in \eqref{2DQCD-TopologicalCoset} however, one must carefully keep track of all topological sectors i.e.\ distinct vacua. (See e.g.\ \cite{Delmastro:2021otj, Cordova:2023jip, Cordova:2024goh}.)

\begin{figure}
        \includegraphics[scale=1.07]{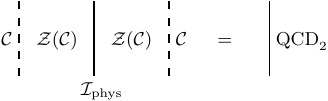} 
        \caption{Three-dimensional construction of $\QCD$.} \label{2DQCD-IntervalCompactification}
\end{figure}

One potential source of vacua of the theory can be seen directly in the ultraviolet from the presence of topological local operators, or equivalently exact (in contrast to emergent) one-form symmetries. Such symmetries arise in particular from the subgroup of the center of the gauge group $G$ which acts trivially on the matter representation $\mathbf{R}.$ As discussed in \cite{Komargodski:2020mxz}, exact one-form symmetries split the theory into distinct \textit{universes}, with no finite energy configurations interpolating between them. A given universe contains, however, its own set of vacua, and it remains meaningful to discuss finite-energy configurations in between such vacua. For simplicity below, we will avoid this phenomenon by focusing on examples with trivial one-form symmetry whose unique exact topological local operator is the identity.

Next, we discuss how to characterize the zero-form symmetries of gapped $\QCD$. These symmetries play a crucial role in our treatment of particle-soliton degeneracy. An insightful approach is to construct $\QCD$ via compactification from three spacetime dimensions on a transverse interval \cite{Komargodski:2020mxz,Delmastro:2022prj, Damia:2024kyt,Cordova:2024goh}. In this viewpoint, $\QCD$ with simply-connected gauge group can be constructed by taking a Chern-Simons theory in bulk:
\begin{equation}
    \mathcal{Z}(\mathcal{C}) \coloneqq \mathrm{Spin}( \mathrm{dim}(\mathbf{R}))_{1} \times G_{-I(\mathbf{R})}~,
\end{equation}
 and setting coset boundary conditions on the left and on the right of the interval. Notice that since we are considering the gapped case, these are \textit{topological boundary conditions} for $\mathcal{Z}(\mathcal{C})$. See Fig. \ref{2DQCD-IntervalCompactification}. An interface $\mathcal{I}_{\mathrm{phys}}$ controlling the RG flow at finite values of the coupling is set in the middle of the interval. The assumption that the IR is described by the topological coset \eqref{gaugedWZW} is then equivalent to the statement that in the far IR the surface $\mathcal{I}_{\mathrm{phys}}$ becomes the transparent identity interface in $\mathcal{Z}(\mathcal{C})$. 
 
 The construction of $\QCD$ sketched above manifests the symmetries. Indeed, given any 3D TQFT, a fusion category $\mathcal{C}$ of line operators is always supported at any topological boundary \cite{Kong:2013aya}.  Viewed purely from two dimensions, $\mathcal{C}$ are topological lines and hence yield the desired symmetries.  Moreover, the 3D construction of $\QCD$ implies that such a fusion category of line operators persists along the whole RG flow and hence can be used to analyze the spectrum.\footnote{One can also see that the quotient of chiral algebras is preserved along the whole RG directly in 2D \cite{Kutasov:1994xq, Delmastro:2021otj,Delmastro:2022prj}.}  In the IR, the surface $\mathcal{I}_{\mathrm{phys}}$ becomes transparent and this symmetry is fully spontaneously broken:
\begin{equation}\label{eqspont}
    \cC  \longrightarrow \mathbf{1}.
\end{equation}
In particular,  the vacua are in one-to-one correspondence with simple objects in $\cC$ \cite{Cordova:2024goh}.  Finally, we remark that, in general, $\cC $ may not be the complete finite symmetry of a given (bosonized) gapped $\QCD$.  However, because of its manifest nature from 3D, $\cC$ can be readily determined in practice and hence its implications for spectra are especially computable.\footnote{We also remark that, following \cite{Cordova:2024goh}, our methods are in principle generalizable to any gauged WZW model described by a conformal embedding in \eqref{bosonic2DQCD} with IR description \eqref{gaugedWZW}.}

\section{Anyon Condensation and Topological Cosets}
Our previous analysis has indicated the central importance of $\cC,$ the fusion category of lines at a topological boundary of a 3D TQFT. To determine $\cC$ directly, we use anyon condensation.  We refer to \cite{Kong:2013aya} for a detailed presentation, and in the following provide a simplified presentation sufficient for our examples. For an introduction to the algebraic theory of anyons, see \cite{Kitaev:2005hzj, Benini:2018reh}. Our conventions follow \cite{Cordova:2024goh}.

\begin{figure} 
        \includegraphics[scale=1.3]{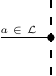} 
        \caption{An anyon $a \in \mathcal{L}$ can end at a topological junction at the topological boundary defined by $\mathcal{L}$.} \label{Anyonending}
\end{figure}

In general, an arbitrary topological boundary condition of a 3D TQFT $\mathcal{T}$ can be described in terms of a non-simple anyon called a \textit{Lagrangian algebra} \cite{Kong:2013aya, Kaidi:2021gbs}:
\begin{align} \label{LagrangianAlgebra}
    \mathcal{L} = \bigoplus_{a \in \mathcal{T}} n_{a} \, a, \quad n_{a} \in \mathbb{N},
\end{align}
where $a$ are simple anyons, and $n_{a}$ positive integers.  For simplicity, we will consider the multiplicity-free case $n_{a} = {0,1}$. A useful interpretation of the Lagrangian algebra is that it dictates the anyons that are allowed to end perpendicularly in a topological junction at the topological boundary described by $\mathcal{L}$. See Fig. \ref{Anyonending}.
Lagrangian algebras can only be composed of bosons. This is, if the topological spin $\theta_{a}\neq 1 \Rightarrow n_{a} = 0$. Moreover, Lagrangian algebras satisfy the quantum dimension constraints $\mathrm{dim}(\mathcal{T}) = \big( \mathrm{dim}(\mathcal{L}) \big)^{2}$, where $\mathrm{dim}(\mathcal{T}) = \sum_{a \in \mathcal{T}} d_{a}^{2}$ and $\mathrm{dim}(\mathcal{L}) = \sum_{a \in \mathcal{T}} n_{a} d_{a}$. In general, it is a non-trivial requirement for a 3D TQFT to have a set of anyons fulfilling these properties, so this already provides a non-trivial tool to characterize topological boundaries. For a complete characterization of Lagrangian algebras, we refer to the aforementioned references.

The line operators at the topological boundary can be found by taking a simple anyon $a$ of the bulk TQFT and moving it to the boundary. Generically, such a simple anyon becomes a non-simple line operator at the boundary. This is encapsulated in the following ``splitting'' rule:
\begin{equation}
    a = \sum_{\alpha} z^{\alpha}_{a} \, \alpha, \quad z^{\alpha}_{a} \in \mathbb{N}, \label{restriction}
\end{equation}
where $\alpha$ stands for the distinct simple line operators of the boundary fusion category. Generically, a simple line operator $\alpha$ may appear with multiplicity in the splitting rule, a fact that is encoded in the non-negative integers $z^{\alpha}_{a}$. Simple anyons in the Lagrangian algebra $\mathcal{L} = \oplus_{a} n_{a} a$ always have a component of the identity line of the boundary theory: $a \rightarrow \, 1 + \cdots$.

The boundary fusion category $\cC$ is constrained to satisfy the following set of consistency conditions:
\begin{itemize}
    \item $a = \sum_{\alpha} z^{\alpha}_{a} \, \alpha \Longrightarrow d_{a} = \sum_{b} z^{\alpha}_{a} \, d_{\alpha}$.

    \item $a = \sum_{\alpha} z^{\alpha}_{a} \, \alpha \Longrightarrow  \bar{a} = \sum_{\alpha} z^{\alpha}_{a} \,  \bar{\alpha}.$

    \item $a \otimes b = \bigoplus_{c}  N_{a,b}^{c} \, c \Longrightarrow \Big( \sum_{\alpha} z^{\alpha}_{a} \, \alpha \Big) \times \Big( \sum_{\beta} z^{\beta}_{b} \, \beta \Big) = \sum_{c, \gamma} \, N_{a,b}^{c} \, z^{\gamma}_{c} \, \gamma.$
\end{itemize}
The boundary fusion category must also satisfy the standard conditions of associativity, existence of a unique identity line, and existence of unique conjugates with a unique way to fuse to the identity line. Often, one can exploit the previous constrains and find the fusion ring of the fusion category exactly. This will be our main practical tool below.

The relation between topological cosets and topological boundary conditions has recently been studied in detail in \cite{Cordova:2024goh}. In short, when we have an embedding of chiral algebras $G_{k} \hookrightarrow \mathcal{Q}_{1}$ with associated gauged WZW model with vanishing central charge and simply-connected $G$, then the Chern-Simons theory $\mathcal{Q}_{1} \times G_{-k}$ admits a topological boundary. (This is the statement that topological cosets belong to the trivial Witt class \cite{davydov2013witt}.) More specifically, the branching rules of the conformal embedding
\begin{equation} \label{branchingrules}
    \chi^{\mathcal{Q}_{1}}_{\Lambda}(q) = \sum_{\lambda} b_{(\Lambda, \lambda)} \, \chi^{G_{k}}_{\lambda}(q), \quad b_{(\Lambda, \lambda)}  \in \mathbb{N},
\end{equation}
with $q$ the modular parameter, induce the existence of a Lagrangian algebra $\mathcal{L} = \bigoplus_{(\Lambda, \lambda)} b_{(\Lambda, \lambda)} (\Lambda, \lambda^{\mathrm{op}})$ defining the corresponding topological boundary of $\mathcal{Q}_{1} \times G_{-k}$. Thus, we can in particular apply this observation to the topological cosets \eqref{2DQCD-TopologicalCoset} pertinent to $\QCD$ and apply the rules of anyon condensation to obtain the fusion categories that are present throughout the RG flow.

As discussed above, in order to avoid the phenomenon of universe splitting in $\QCD$, we may consider global forms of the gauge group that have trivial center. To address this in the 3D construction, we follow \cite{Moore:1989yh,Hori:1994nc, Cordova:2023jip} and gauge the ``common center'' one-form symmetry in $\mathrm{Spin}( \mathrm{dim}(\mathbf{R}))_{1} \times G_{-I(\mathbf{R})}$. After this is done, no exact one-form symmetry given by the center of the gauge group remains, and all the topological local operators in the IR correspond to vacua of a single universe. Notice that a gapped boundary for the gauged theory $(\mathrm{Spin}( \mathrm{dim}(\mathbf{R}))_{1} \times G_{-I(\mathbf{R})})/Z$ (with $Z$ the ``common center'') is guaranteed to exist given the gapped boundary determined by \eqref{branchingrules}, since the gauging by $Z$ is just a topological manipulation. In practice, a Lagrangian algebra in $(\mathrm{Spin}( \mathrm{dim}(\mathbf{R}))_{1} \times G_{-I(\mathbf{R})})/Z$ is easily found by direct examination of the TQFT.

Anyon condensation also allows us to make statements regarding the expectation values in the different vacua of the local operators that do not decouple in the IR fixed point. Indeed, local operators in the 2D theory in the IR are constructed from stretching anyons between the two ends of the interval:
\begin{align} \label{toplocaloperatorviaLagrangian}
\begin{gathered}
    \includegraphics[scale=1.3]{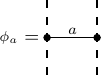} 
\end{gathered}
\end{align}
Thus, Lagrangian algebras determine the set of topological local operators, or vacua in the IR of $\QCD$. Below, the topological OPE of these local operators can be obtained exploiting commutativity, associativity, consistency of the OPE with the local operators allowed by the Lagrangian algebra, and existence of an idempotent complete basis $\{ \mathbf{v}_{i} \}$ of local operators that are in one-to-one correspondence with the clustering vacua of the theory. In turn, determining the $\phi_{a}$ in terms of the idempotent complete basis allows us to calculate the vacuum expectation value of each local operator $\phi_{a}.$

\section{Representation Theory}
Having identified a collection of finite non-invertible symmetry in gapped $\QCD$ theories, we now consider their implications for particle spectra. Recent work \cite{Cordova:2024vsq,Cordova:2024iti} has shown that non-invertible symmetry in a QFT can enforce mass degeneracies. An especially novel feature of such degeneracies is they can be between stable particle and soliton excitations. In this section we briefly review the representation theory that governs such degeneracies, following and deferring detailed discussion to \cite{Cordova:2024iti}.

We first consider a general 2D bosonic QFT before returning to the specific context of $\QCD$.  In general, the finite symmetry is described by a fusion category $\cC$ \cite{Frohlich:2009gb,Carqueville:2012Dk,Bhardwaj:2017xup,Chang:2018iay} and the phase of the symmetry is described by a $\cC$-module category $\cM$ \cite{Bhardwaj:2017xup,Thorngren:2019iar,Huang:2021zvu}. In a gapped theory quantized on $\R$, the simple objects of $\cM$ correspond to the clustering grounding states of the theory \cite{Komargodski:2020mxz,Cordova:2024vsq}.

In a gapped 2D QFT quantized on $\R$, the action of the finite non-invertible symmetry on the state space is through an algebra called the ``strip algebra''. This algebra depends both on the symmetry $(\cC)$ and its phase $(\cM)$ and is denoted $\Str_{\cC}(\cM)$.  It is the representation theory of the strip algebra that governs symmetry enforced degeneracies of the theory. $\Str_{\cC}(\cM)$ is a $C^*$-weak Hopf algebra, which means that its representations can act on multi-particle states (tensor products), have charge conjugates (duals), and are compatible with the unitary structure on the Hilbert space of states. The representation theory of such algebras is well studied and admits techniques for analysis similar to those used in the unitary representation theory of finite groups. In fact, in the special case of a symmetry described by a finite group $H$ in its unbroken phase, $\Str_\cC(\cM) = \C[H]$, the group algebra of $H$, and hence recovers the familiar representation theory of finite groups.

In analyzing the degeneracies of gapped $\QCD$ theories, one can leverage that the representation category $\Rep(\Str_\cC(\cM))$ also admits a more abstract description,
\begin{equation}
    \Rep(\Str_{\cC}(\cM)) \simeq \cC^*_\cM.
\end{equation}
Here $\cC^*_\cM$ is called the ``dual category'' of $\cC$ with respect to $\cM$ and is physically the dual symmetry obtained by performing a generalized gauging of $\cC$ associated to $\cM$ \cite{Bhardwaj:2017xup,Diatlyk:2023fwf}. This category naturally acts on $\cM$. Taking $\alpha$ to be a simple line, its corresponding irreducible representation is the boundary junction vector space
\begin{equation}\label{eq:reps_and_simple_lines}
    V_\alpha = \bigoplus_{m,n}\Hom(m \otimes \alpha,n),
\end{equation}
where $m,n$ are the clustering ground states of the theory. This presentation of the representation is especially useful since it makes clear how many particles (solitons) there are above (between) each vacuum in it.

To compute the action of $\Str_{\cC}(\cM)$ on $V_\alpha$ in this approach, it is necessary to know the complete $(\cC$-$\cC^*_\cM)$-bimodule category structure on $\cM$. This data is generically difficult to compute. Fortunately, equation \eqref{eq:reps_and_simple_lines} contains a significant amount of information about the representation and is determined by far less data. The dimensions of each summand is given by the module category fusion coefficients
\begin{equation}
    m \otimes \alpha= \bigoplus_{n}\tilde{N}_{m,\alpha}^n n, \quad \tilde{N}_{m,\alpha}^n \in \mathbb{N}.
\end{equation}
The fusion coefficients can be conveniently encoded in a quiver (i.e.\ a directed graph):
\begin{itemize}
    \item For each clustering ground state $m \in \cM$ draw a node of the quiver.
    \item Add $\tilde{N}_{m, \alpha}^n$ directed arrows from node $m$ to node $n$.
\end{itemize}
This provides a convenient graphical presentation of the module fusion coefficients. 

We can apply this discussion to analyze symmetry-enforced degeneracies of gapped $\QCD$ theories. As discussed in Section \ref{sec:background}, the relevant fusion category symmetry is the category of lines of the coset boundary condition of the 3D coset TQFT, $\cC$. The assumed symmetry breaking pattern \eqref{eqspont}, implies that the phase of the symmetry is given by $\cM = \cC$ viewed as a $\cC$-module category. Therefore, we must compute $\cC_\cC^*$, which is just the category itself \cite{ReferenceKey},
\begin{equation}
    \cC^*_\cC \simeq \cC.
\end{equation}
To compute the quivers describing the allowed symmetry enforced degeneracies one therefore only needs to know the fusion rules for $\cC$
\begin{equation}
    \tilde{N} = N, \quad \alpha \otimes \beta = \bigoplus_{\gamma}N_{\alpha,\beta}^\gamma \gamma.
\end{equation}

Before continuing to examples, a comment regarding one-form symmetry is in order. The mixing of particle and soliton states by non-invertible symmetry is particularly powerful because it can require the existence of particle states in a theory. This is unlike the case of group-like symmetry where the realization of a particular representation is a dynamical rather than kinematic question. This follows from the fact that a theory has no non-trivial one-form symmetry if and only if the quiver describing the complete collection of stable solitons is connected. We provide a short argument for this in Appendix \ref{app:one_form_disc}, where we also discuss the precise implications of this on the possible realized representations. As will be demonstrated in the following examples, this additional information is often enough to require the existence of some stable particles or solitons from purely kinematic considerations. 

\section{Examples of Particle-Soliton Degeneracy} 
We now consider examples of this analysis in gapped $\QCD$ theories. In doing so we demonstrate the necessary existence of a collection of stable particle and/or soliton states in each theory. Details of the computations are included in Appendix \ref{app:example_calcs}, including the data of the boundary condensation maps and allowed quivers for each of the following theories.

\smallskip
\noindent $\bm{SO(3)} + \psi_{\bm{5}}$: Our first example is $SO(3)$ gauge theory with Majorana fermions in the five. The bosonized theory has the 3D coset TQFT
\begin{equation}\label{eq:SO(3)_coset_TQFT}
    \frac{Spin(5)_1 \times SU(2)_{-10}}{\Z_2} = \cZ(\cC).
\end{equation}
Its coset boundary condition has 3 lines $\{1,v,A\}$ with the non-trivial fusions
\begin{equation}\label{eq:SO(3)_fusion_table}
\begin{tabular}{ |c|c|c| } 
\hline
$\times$  & $v$ & $A$ \\
\hline
$v$ & $1$ & $A$ \\ 
\hline
$A$ & $A$ & $1 + v + 2A$ \\ 
\hline
\end{tabular}~.
\end{equation}
The $\QCD$ theory therefore has $3$ vacua and $3$ possible irreducible multiplets. Because the theory has no one-form symmetry along the flow, the stable solitons and particles must realize a representation furnishing a connected quiver. All such quivers contain
\begin{equation}\label{eq:SO(3)_min_quiver}
\begin{tikzcd}
	1 & A & v
	\arrow[tail reversed, from=1-1, to=1-2]
	\arrow[from=1-2, to=1-2, loop, in=60, out=120, distance=5mm]
	\arrow[from=1-2, to=1-2, loop, in=55, out=125, distance=10mm]
	\arrow[tail reversed, from=1-2, to=1-3]
\end{tikzcd}~,
\end{equation}
as a sub-quiver. It follows that this theory must contain at least two stable particles over the $\ket{\Omega_A}$ vacuum and two soliton-anti-soliton pairs. Furthermore, this quiver is realized by an irreducible representation of the strip algebra and therefore the excitations all have equal mass.

\smallskip
\noindent $\bm{Spin(9)} + \psi_{\sigma}$: Our next example is $Spin(9)$ gauge theory with Majorana fermions in the spinorial. The bosonized theory has the 3D coset TQFT
\begin{equation}\label{eq:Spin(9)_coset_TQFT}
    Spin(16)_1 \times Spin(9)_{-2} = \cZ(\cC).
\end{equation}
The coset boundary condition has $6$ lines $\{1,s,v,c,A,B\}$ with non-trivial fusions
\begin{equation}\label{eq:Spin(9)_fusion_table}
\begin{tabular}{ |c|c|c|c|c|c| } 
\hline
$\times$  & $ {s}$ & $ {v}$ & $ {c}$ & $A$ & $B$ \\
\hline
$ {s}$ & $1$ & $ {c}$  & $ {v}$  & $A$  & $B$ \\ 
\hline
$ {v}$ & $ {c}$ & $1$  & $ {s}$  & $B$  & $A$ \\ 
\hline
$ {c}$ & $ {v}$ & $ {s}$  & $1$  & $B$  & $A$ \\ 
\hline
$A$ & $A$ & $B$  & $B$  & $1 +  {s} + A$  & $ {v} +  {c} + B$ \\ 
\hline
$B$ & $B$ & $A$  & $A$  & $ {v} +  {c} + B$  & $1 +  {s} + A$ \\ 
\hline
\end{tabular}~.
\end{equation}
The theory therefore has $6$ vacua and $6$ possible irreducible representations. This theory also does not have one-form symmetry along the flow and so its stable excitations must realize representations assembling into a connected quiver. All such quivers contain
\begin{equation}\label{eq:Spin(9)_min_quiver}
\begin{tikzcd}
	& v & 1 \\
	c & A & B & s
	\arrow[tail reversed, from=2-2, to=2-3]
\end{tikzcd}~,
\end{equation}
as a sub-quiver. Therefore this theory must contain a soliton-anti-soliton pair of equal mass. Moreover, this pair necessarily exists as a member of one of three possible minimal connected sub-quivers\footnote{\label{ft:minimal quiver}Meaning that every connected quiver of stable solitons must contain one of these as a sub-quiver.}
\begin{equation}\label{eq:spin_quiv_1}
\begin{gathered}
\begin{tikzcd}
	& 1 & v \\
	c &&& s \\
	& A & B
	\arrow[tail reversed, from=1-2, to=3-3]
	\arrow[tail reversed, from=2-1, to=3-2]
	\arrow[tail reversed, from=3-2, to=1-3]
	\arrow[tail reversed, from=3-2, to=3-3]
	\arrow[tail reversed, from=3-3, to=2-4]
\end{tikzcd}\end{gathered}~,
\end{equation}
\begin{equation}\label{eq:spin_quiv_2}
\begin{tikzcd}
	& 1 & v \\
	c &&& s \\
	& A & B
	\arrow[color={rgb,255:red,194;green,20;blue,38}, tail reversed, from=1-2, to=3-2]
	\arrow[tail reversed, from=1-3, to=2-4]
	\arrow[color={rgb,255:red,194;green,20;blue,38}, tail reversed, from=1-3, to=3-3]
	\arrow[tail reversed, from=2-1, to=1-2]
	\arrow[color={rgb,255:red,194;green,20;blue,38}, tail reversed, from=2-1, to=3-3]
	\arrow[color={rgb,255:red,194;green,20;blue,38}, tail reversed, from=3-2, to=2-4]
	\arrow[tail reversed, from=3-2, to=3-3]
\end{tikzcd}~,
\end{equation}
\begin{equation}\label{eq:spin_quiv_3}
\begin{tikzcd}
	& 1 & v \\
	c &&& s \\
	& A & B
	\arrow[tail reversed, from=1-2, to=1-3]
	\arrow[color={rgb,255:red,194;green,20;blue,38}, tail reversed, from=1-2, to=3-2]
	\arrow[color={rgb,255:red,194;green,20;blue,38}, tail reversed, from=1-3, to=3-3]
	\arrow[tail reversed, from=2-1, to=2-4]
	\arrow[color={rgb,255:red,194;green,20;blue,38}, tail reversed, from=2-1, to=3-3]
	\arrow[color={rgb,255:red,194;green,20;blue,38}, tail reversed, from=3-2, to=2-4]
	\arrow[tail reversed, from=3-2, to=3-3]
\end{tikzcd}~,
\end{equation}
where colors are used to distinguish the sub-quivers of irreducible representations comprising a reducible representation. One such quiver must be realized, but symmetry considerations alone do not determine which. This instead is a question of dynamics. Note that in the later two cases, since the representations are reducible, not all solitons are required to have degenerate masses. Rather, this allows for two multiplets of stable solitons, each having a different mass.

\smallskip
\noindent $\bm{PSU(4)} + \psi_{\bm{15}}$: Our final example is $PSU(4)$ gauge theory with Majorana fermions in the fifteen. The bosonized theory has the 3D coset TQFT
\begin{equation}\label{eq:PSU(4)_coset_TQFT}
    \frac{Spin(15)_1 \times SU(4)_{-4}}{\Z_4} = \cZ(\cC).
\end{equation}
The coset boundary condition has $4$ lines $\{1,v,A,B\}$ with non-trivial fusions
\begin{equation}\label{eq:PSU(4)_fusion_table}
\begin{tabular}{ |c|c|c|c| } 
\hline
$\times$  & $v$ & $A$ & $B$ \\
\hline
$v$ & $1$ & $B$ & $A$ \\ 
\hline
$A$ & $B$ & $1 + A + B$ & $v + A + B$ \\ 
\hline
$B$ & $A$ & $v + A + B$ &  $1 + A + B$ \\ 
\hline
\end{tabular}~.
\end{equation}
The theory therefore has $4$ vacua and $4$ possible irreducible multiplets. As in the prior examples, this theory has no one-form symmetry along the flow and so its stable particles and solitons must furnish representations that combine to give a connected quiver. All such quivers contain 
\begin{equation}\label{eq:PSU(4)_min_quiver}
\begin{tikzcd}
	1 & A & B & v
	\arrow[from=1-2, to=1-2, loop, in=60, out=120, distance=5mm]
	\arrow[tail reversed, from=1-2, to=1-3]
	\arrow[from=1-3, to=1-3, loop, in=60, out=120, distance=5mm]
\end{tikzcd}~,
\end{equation}
as a sub-quiver. Therefore the theory contains at least two stable particles, one above $\ket{\Omega_A}$ and the other above $\ket{\Omega_B}$, and a soliton-anti-soliton pair between these vacua.

These excitations are necessarily members of one of two possible minimal connected sub-quivers,
\begin{equation}\begin{gathered}
\begin{tikzcd}
	1 && B \\
	\\
	A && v
	\arrow[tail reversed, from=1-1, to=3-1]
	\arrow[from=1-3, to=1-3, loop, in=60, out=120, distance=5mm]
	\arrow[tail reversed, from=1-3, to=3-3]
	\arrow[tail reversed, from=3-1, to=1-3]
	\arrow[from=3-1, to=3-1, loop, in=240, out=300, distance=5mm]
\end{tikzcd}, \quad
\begin{tikzcd}
	1 && B \\
	\\
	A && v
	\arrow[tail reversed, from=1-1, to=1-3]
	\arrow[from=1-3, to=1-3, loop, in=60, out=120, distance=5mm]
	\arrow[tail reversed, from=3-1, to=1-3]
	\arrow[from=3-1, to=3-1, loop, in=240, out=300, distance=5mm]
	\arrow[tail reversed, from=3-1, to=3-3]
\end{tikzcd}\end{gathered}~.\end{equation} 
Again, which sub-quiver is realized by the theory is not determined by symmetry, but by dynamics. Both  quivers correspond to irreducible representations and therefore all stable particles and solitons labeled by them will have degenerate mass.

\let\oldaddcontentsline\addcontentsline
\renewcommand{\addcontentsline}[3]{}
\section*{Acknowledgments}
\let\addcontentsline\oldaddcontentsline

CC, DGS, and NH acknowledge support from the Simons Collaboration on Global Categorical Symmetries, the US Department of Energy Grant 5-29073, and the Sloan Foundation.

\appendix

\section{One-form Symmetry and Stable Solitons.}\label{app:one_form_disc}

In this section, we clarify the relationship between one-form symmetry and the existence of stable solitons used in the main text to demonstrate the necessary existence of stable solitons and particles in the theories considered. 

Let $\cT$ denote a gapped 2D QFT realizing a spontaneously broken phase ($\cM$) of some finite symmetry ($\cC$). The Hilbert space quantized on $\R$ then decomposes into sectors
\begin{equation}
    \cH = \bigoplus_{m,n} \cH_{m,n},
\end{equation}
where $\cH_{m,n}$ is the subspace of states that resemble the clustering ground state $\ket{\Omega_m}$ as $x \to -\infty$ and the clustering ground state $\ket{\Omega_n}$ as $x \to \infty$. If $\cT$ has non-trivial one-form symmetry, then it follows that some sector $\cH_{m,n}$ must be zero, meaning there are no finite tension domain walls between the vacua \cite{Komargodski:2020mxz}. 

It's natural to ask if the converse is true. That is, does the non-existence of finite energy domain walls between two vacua imply the existence of a non-trivial topological local operator (one-form symmetry)? A quick argument for this is the following. Suppose that there are two vacua such that $\cH_{m,n}$ is zero. Then it's partition function must vanish
\begin{equation}
    \tr\left(e^{-2\pi tH_{m,n}}\right) = 0.
\end{equation}
If we consider $\R$ as the infinite length limit of a finite interval with boundary conditions, then this is a limit of
\begin{equation}
    \tr\left(e^{-2\pi tH_{m,n}(L)}\right) \xrightarrow[L \to \infty]{} 0,
\end{equation}
with $H_{m,n}(L)$ the Hamiltonian of a finite interval of length $L$. The finite $L$ expression can alternatively be computed using states corresponding to a radial quantization of radius $t$,
\begin{equation}
    \tr\left(e^{-2\pi tH_{m,n}(L)}\right) = \braket{\bm v_m|e^{-L\tilde{H}(t)}|\bm v_n}.
\end{equation}
The right hand expression is computed in a radial quantization $\cH(S^1_t)$, with $\tilde{H}(t)$ the radial Hamiltonian and $\ket{\bm v_m}$ the boundary state corresponding to the boundary condition $m$. Computing this in the $L \to \infty$ limit projects onto contributions from the ground states on the circle
\begin{equation}
    \braket{\bm v_m|e^{-L\tilde{H}(t)}|\bm v_n} \xrightarrow[L \to \infty]{} \sum_i\braket{\bm v_m|0_i}\braket{0_i|\bm v_n},
\end{equation}
with $\ket{0_i}$ the ground states in $\mathcal{H}(S^1_t)$.

We claim that the condition $\cH_{m,n} = 0$ implies that there are multiple ground states in this sum. Towards a contradiction, assume there is only one. Then the sequence of above equalities implies
\begin{equation}
    \braket{\bm v_m|0}\braket{0|\bm v_n} = 0.
\end{equation}
Up to relabeling suppose $\braket{\bm v_m|0} = 0$. Now the previous calculations with $m = n$ show that
\begin{equation}
    \tr\left(e^{-2\pi tH_{m,m}}\right) = \braket{\bm v_m|0}\braket{0|\bm v_m},
\end{equation}
which is then zero,
\begin{equation}
    \tr\left(e^{-2\pi tH_{m,m}}\right) = 0.
\end{equation}
Since $e^{-2\pi tH_{m,m}}$ is a positive operator, this implies that $\cH_{m,m} = 0$. But this cannot be so, since we require that $\cH_{m,m}$ contains a non-zero ground state $\ket{\Omega_m}$. Therefore the radial Hilbert space $\cH(S^1_t)$ must have degenerate ground states. This is true for every radius.

Let us suppose that as $t \to 0$, we recover the radial Hilbert space of the UV CFT of $\cT$,
\begin{equation}
    \cH(S^1_t) \xrightarrow[t \to 0]{} \cH(S^1)_{UV}.
\end{equation}
These arguments then imply that the UV CFT has multiple ground states. By the state-operator correspondence it therefore has multiple topological local operators, that is, non-trivial one-form symmetry. Since these operators can factor the CFT in the UV, they therefore exist in $\cT$.

Continuing on, assuming the absence of one-form symmetry, we now ask what is implied about the stable soliton spectrum of the theory? The stable soliton content can be represented by a quiver, with nodes labeled by clustering ground states $m$ and there a directed arrow from $m$ to $n$ for every stable soliton in $\cH_{m,n}$. The absence of one-form symmetry requires this quiver be connected. To see why this is true, consider a generic state $\ket{\psi} \in \cH_{m,n}$. At late times it must decay into a collection of stable solitons and particles. The vacuum labels of neighboring solitons must agree and so any out-going summand defines a path from $m$ to $n$ in the quiver. Therefore the quiver is connected.

Finally, we can ask what this implies about the representations of $\Str_\cC(\cM)$ that stable solitons can furnish. Under the same working assumptions, suppose that $\cT$ has stable solitons realizing $N$ irreducible representations $\{R_i\}_{i}$, $i = 1, \dots, N$. The condition is that the quiver associated to the representation $\oplus_i R_i$ is connected, which is just a restatement of the prior paragraph. Any sub-quiver common to all such quivers must therefore be realized in a theory without one-form symmetry. As demonstrated in the main text, non-trivial sub-quivers can exist in practice, providing a purely kinematic explanation of the existence of some solitons and particles in a theory.

For the study of particle degeneracies, it is important to note that when the stable solitons realize an irreducible representation of $\Str_\cC(\cM)$, they are necessarily of degenerate mass \cite{Cordova:2024vsq}. In contrast, if the stable solitons instead realize a reducible representation, $\bigoplus_i R_i$, solitons in different subrepresentations can have different masses.

\section{ Quiver Calculations}\label{app:example_calcs}
In this section, we show how to calculate the quivers of the gapped $\QCD$ theories discussed in the main text. Recall we are interested in topological cosets where for simply-connected gauge group the bulk 3D TQFT is given by 
\begin{equation}
    \mathrm{Spin}( \mathrm{dim}(\mathbf{R}))_{1} \times G_{-I(\mathbf{R})}.
\end{equation}
The line operators in this 3D TQFT will be labeled as $(\Lambda, \lambda)$, with $\Lambda$ an integrable representation of $\mathrm{Spin}( \mathrm{dim}(\mathbf{R}))_{1}$ and $\lambda$ an integrable representation of $G_{I(\mathbf{R})}$, with the negative level implicit in the notation. When $\mathrm{dim}(\mathbf{R})$ is even, we denote the four integrable representation of $\mathrm{Spin}( \mathrm{dim}(\mathbf{R}))_{1}$ as $\{\mathbf{1}, \mathbf{v}, \mathbf{s}, \mathbf{c} \}$ with $\mathbf{1}$ labeling the identity, $\mathbf{v}$ the vectorial, and $\mathbf{s}$ and $\mathbf{c}$ the spinorials. When $\mathrm{dim}(\mathbf{R})$ is odd, we denote the three integrable representations of $\mathrm{Spin}( \mathrm{dim}(\mathbf{R}))_{1}$ as $\{\mathbf{1}, \mathbf{v}, \mathbf{s}\}$ with $\mathbf{v}, \mathbf{s}$ the vectorial and spinorial representations respectively. When the gauge group is not simply-connected, we denote the lines of the bulk TQFT $(\mathrm{Spin}( \mathrm{dim}(\mathbf{R}))_{1} \times G_{-I(\mathbf{R})})/Z$ with $Z$ the ``common center'' in terms of their representatives in the simply-connected case before gauging $Z$.

\smallskip
\noindent $\bm{SO(3)}+\psi_{\bm5}$.
For the bosonized theory, the 3D coset TQFT is
\begin{equation}\tag{\ref{eq:SO(3)_coset_TQFT}}
    \frac{Spin(5)_1 \times SU(2)_{-10}}{\Z_2} = \cZ(\cC).
\end{equation}
This theory has $10$ lines while its coset boundary condition has $3$, $\{1,v,A\}$. The condensation maps from the 3D theory to the coset boundary are
\begin{equation}
\begin{gathered}
\begin{tabular}{ |c|c| } 
\hline
$a$ & $p(a)$ \\
\hline
$(\mathbf{1}, \mathbf{1})$ & $1$ \\
\hline
$(\mathbf{1}, \mathbf{11})$ & $v$ \\
\hline
$(\mathbf{1}, \mathbf{3})$  & $A$ \\
\hline
$(\mathbf{1}, \mathbf{9})$ &  $A$ \\
\hline
$(\mathbf{1}, \mathbf{5})$ & $v + A$ \\
\hline
\end{tabular} \quad 
\begin{tabular}{ |c|c| } 
\hline
$a$ & $p(a)$ \\
\hline
$(\mathbf{1}, \mathbf{7})$ & $1 + A$\\
\hline
$(\mathbf{s}, \mathbf{2})$ & $A$ \\
\hline
$(\mathbf{s}, \mathbf{4})$ & $1+v+A$ \\
\hline
$(\mathbf{s}, \mathbf{6})_{1}$ & $A$ \\
\hline
$(\mathbf{s}, \mathbf{6})_{2}$ & $A$ \\
\hline
\end{tabular}
\end{gathered}~.
\end{equation}
The boundary lines have non-trivial fusions
\begin{equation}\tag{\ref{eq:SO(3)_fusion_table}}
\begin{tabular}{ |c|c|c| } 
\hline
$\times$  & $v$ & $A$ \\
\hline
$v$ & $1$ & $A$ \\ 
\hline
$A$ & $A$ & $1 + v + 2A$ \\ 
\hline
\end{tabular}~.
\end{equation}
Because the coset boundary has $3$ lines, there are therefore $3$ irreducible representations of the strip algebra, with one labeled by each line.

Recall from the main text that the module category fusion coefficients are simply the fusion coefficients which are computed from \eqref{eq:SO(3)_fusion_table}, 
\begin{equation}
\begin{gathered}
    N_{1,1}^1 = N_{v,1}^v = N_{A,1}^A = 1, \\
    N_{1,v}^{v} = N_{v,v}^1 = N_{A,v}^A = 1, \\
    N_{1,A}^A = N_{v,A}^A = N_{A,A}^{1} = N_{A,A}^v = 1, \; N_{A,A}^A = 2.
\end{gathered}
\end{equation}
Therefore the quivers realized by $1$, $v$, and $A$ are
\begin{equation} 1 : \quad 
\begin{tikzcd}
	1 & A & v
	\arrow[from=1-1, to=1-1, loop, in=60, out=120, distance=5mm]
	\arrow[from=1-2, to=1-2, loop, in=60, out=120, distance=5mm]
	\arrow[from=1-3, to=1-3, loop, in=60, out=120, distance=5mm]
\end{tikzcd}~,
\end{equation} 

\begin{equation}
v: \quad
\begin{tikzcd}
	1 & A & v
	\arrow[shift right, curve={height=12pt}, tail reversed, from=1-1, to=1-3]
	\arrow[from=1-2, to=1-2, loop, in=60, out=120, distance=5mm]
\end{tikzcd}~,
\end{equation} 

\begin{equation}\label{eq:app_so(3)_min_quiv}
A: \quad
\begin{tikzcd}
	1 & A & v
	\arrow[tail reversed, from=1-1, to=1-2]
	\arrow[from=1-2, to=1-2, loop, in=55, out=125, distance=10mm]
	\arrow[from=1-2, to=1-2, loop, in=60, out=120, distance=5mm]
	\arrow[tail reversed, from=1-2, to=1-3]
\end{tikzcd}~.
\end{equation} 
The single minimal connected sub-quiver\footnote{See footnote \ref{ft:minimal quiver}.} is clearly \eqref{eq:app_so(3)_min_quiv}. Therefore this must be realized as a sub-quiver in the $\QCD$ theory, requiring the particle-soliton degeneracies discussed in the bulk text.

\smallskip
\noindent $\bm{Spin(9)}+\psi_{\sigma}$.
For the bosonized theory, the 3D coset TQFT is
\begin{equation}\tag{\ref{eq:Spin(9)_coset_TQFT}}
    Spin(16)_1 \times Spin(9)_{-2} = \cZ(\cC).
\end{equation}
This theory has $32$ lines while its coset boundary condition has $6$ lines, $\{1,s,v,c,A,B\}$.
The condensation maps from the 3D theory to the coset boundary are:
\begin{equation}
\begin{gathered}
\begin{tabular}{|L|L|} 
\hline
a & p(a) \\
\hline
(\mathbf{1},\mathbf{1}) & 1 \\ 
\hline
(\mathbf{1},\mathbf{44}) & {s} \\ 
\hline
(\mathbf{1},\mathbf{16}) & {v} + B \\ 
\hline
(\mathbf{1},\mathbf{128}) & {c} + B \\ 
\hline
(\mathbf{1},\mathbf{126}) & A \\
\hline
(\mathbf{1},\mathbf{84}) & 1 + {s} \\
\hline
(\mathbf{1},\mathbf{36}) & A \\
\hline
(\mathbf{1},\mathbf{9}) & A \\
\hline
\end{tabular} \quad
\begin{tabular}{|L|L|} 
\hline
a & p(a) \\
\hline
(\mathbf{v},\mathbf{1}) & {v} \\ 
\hline
(\mathbf{v},\mathbf{44}) & {c} \\ 
\hline
(\mathbf{v},\mathbf{16}) & 1 + A \\ 
\hline
(\mathbf{v},\mathbf{128}) & {s} + A \\ 
\hline
(\mathbf{v},\mathbf{126}) & B \\
\hline
(\mathbf{v},\mathbf{84}) & {v} + {c} \\
\hline
(\mathbf{v},\mathbf{36}) & B \\
\hline
(\mathbf{v},\mathbf{9}) & B \\
\hline
\end{tabular} \\
\begin{tabular}{|L|L|} 
\hline
a & p(a) \\
\hline
(\mathbf{s},\mathbf{1}) & {s} \\ 
\hline
(\mathbf{s},\mathbf{44}) & 1 \\ 
\hline
(\mathbf{s},\mathbf{16}) & {c} + B \\
\hline
(\mathbf{s},\mathbf{128}) & {v} + B \\
\hline
(\mathbf{s},\mathbf{126}) & A \\
\hline
(\mathbf{s},\mathbf{84}) & 1 + {s} \\
\hline
(\mathbf{s},\mathbf{36}) & A \\
\hline
(\mathbf{s},\mathbf{9}) & A \\
\hline
\end{tabular} \quad
\begin{tabular}{|L|L|} 
\hline
a & p(a) \\
\hline
(\mathbf{c},\mathbf{1}) & {c} \\ 
\hline
(\mathbf{c},\mathbf{44}) & {v} \\ 
\hline
(\mathbf{c},\mathbf{16}) & {s} + A \\
\hline
(\mathbf{c},\mathbf{128}) & 1 + A \\ 
\hline
(\mathbf{c},\mathbf{126}) & B \\
\hline
(\mathbf{c},\mathbf{84}) & {v} + {c} \\
\hline
(\mathbf{c},\mathbf{36}) & B \\
\hline
(\mathbf{c},\mathbf{9}) & B \\ 
\hline
\end{tabular}
\end{gathered}~.
\end{equation}
The boundary lines have non-trivial fusions
\begin{equation}\tag{\ref{eq:Spin(9)_fusion_table}}
\begin{tabular}{ |c|c|c|c|c|c| } 
\hline
$\times$  & $ {s}$ & $ {v}$ & $ {c}$ & $A$ & $B$ \\
\hline
$ {s}$ & $1$ & $ {c}$  & $ {v}$  & $A$  & $B$ \\ 
\hline
$ {v}$ & $ {c}$ & $1$  & $ {s}$  & $B$  & $A$ \\ 
\hline
$ {c}$ & $ {v}$ & $ {s}$  & $1$  & $B$  & $A$ \\ 
\hline
$A$ & $A$ & $B$  & $B$  & $1 +  {s} + A$  & $ {v} +  {c} + B$ \\ 
\hline
$B$ & $B$ & $A$  & $A$  & $ {v} +  {c} + B$  & $1 +  {s} + A$ \\ 
\hline
\end{tabular}~.
\end{equation}
There are therefore $6$ distinct irreducible representations of the strip algebra, labeled by each boundary line in $\cC$. The module fusion coefficients (and hence quivers) again follow from \eqref{eq:Spin(9)_fusion_table}.

For example, consider the line $A$. The relevant non-zero fusion coefficients in $\cC$ are 
\begin{equation}
\begin{gathered}
    N_{1,A}^A = N_{v,A}^B= N_{c,A}^B = N_{s,A}^A = 1, \\
    \begin{alignedat}{5}
        N_{A,A}^{1} &= N_{A,A}^s &&= N_{A,A}^A &&= 1, \\
        N_{B,A}^{v} &= N_{B,A}^c &&= N_{B,A}^B &&= 1.
    \end{alignedat}
\end{gathered}
\end{equation}
These realize the quiver
\begin{equation}
A: \quad
\begin{tikzcd}
	& 1 & v \\
	c &&& s \\
	& A & B
	\arrow[tail reversed, from=1-2, to=3-2]
	\arrow[tail reversed, from=1-3, to=3-3]
	\arrow[tail reversed, from=3-2, to=2-4]
	\arrow[from=3-2, to=3-2, loop, in=240, out=300, distance=5mm]
	\arrow[tail reversed, from=3-3, to=2-1]
	\arrow[from=3-3, to=3-3, loop, in=240, out=300, distance=5mm]
\end{tikzcd}~.
\end{equation}
The remaining quivers are
\begin{equation}
1 : \quad
\begin{tikzcd}
	& 1 & v \\
	c &&& s \\
	& A & B
	\arrow[from=1-2, to=1-2, loop, in=60, out=120, distance=5mm]
	\arrow[from=1-3, to=1-3, loop, in=60, out=120, distance=5mm]
	\arrow[from=2-1, to=2-1, loop, in=60, out=120, distance=5mm]
	\arrow[from=2-4, to=2-4, loop, in=60, out=120, distance=5mm]
	\arrow[from=3-2, to=3-2, loop, in=60, out=120, distance=5mm]
	\arrow[from=3-3, to=3-3, loop, in=60, out=120, distance=5mm]
\end{tikzcd}~,
\end{equation}

\begin{equation}
s: \quad
\begin{tikzcd}
	& 1 & v \\
	c &&& s \\
	& A & B
	\arrow[tail reversed, from=1-2, to=2-4]
	\arrow[tail reversed, from=2-1, to=1-3]
	\arrow[from=3-2, to=3-2, loop, in=60, out=120, distance=5mm]
	\arrow[from=3-3, to=3-3, loop, in=60, out=120, distance=5mm]
\end{tikzcd}~,
\end{equation}

\begin{equation}
v: \quad
\begin{tikzcd}
	& 1 & v \\
	c &&& s \\
	& A & B
	\arrow[tail reversed, from=1-2, to=1-3]
	\arrow[tail reversed, from=2-1, to=2-4]
	\arrow[tail reversed, from=3-2, to=3-3]
\end{tikzcd}~,
\end{equation}

\begin{equation}
c: \quad
\begin{tikzcd}
	& 1 & v \\
	c &&& s \\
	& A & B
	\arrow[tail reversed, from=1-3, to=2-4]
	\arrow[tail reversed, from=2-1, to=1-2]
	\arrow[tail reversed, from=3-2, to=3-3]
\end{tikzcd}~,
\end{equation}

\begin{equation}
B: \quad
\begin{tikzcd}
	& 1 & v \\
	c &&& s \\
	& A & B
	\arrow[tail reversed, from=1-2, to=3-3]
	\arrow[tail reversed, from=2-1, to=3-2]
	\arrow[tail reversed, from=3-2, to=1-3]
	\arrow[tail reversed, from=3-2, to=3-3]
	\arrow[tail reversed, from=3-3, to=2-4]
\end{tikzcd}~.
\end{equation}

We would now like to find the minimal connected sub-quivers. Since the quiver labeled by $B$ is connected it is clearly one. By analyzing sums of quivers directly, one finds that there are two more minimal connected subquivers corresponding to reducible representations:

\begin{equation}
\begin{gathered}
c + \textcolor{VividBurgundy}{A}: 
\begin{tikzcd}
	& 1 & v \\
	c &&& s \\
	& A & B
	\arrow[color={rgb,255:red,194;green,20;blue,38}, tail reversed, from=1-2, to=3-2]
	\arrow[tail reversed, from=1-3, to=2-4]
	\arrow[color={rgb,255:red,194;green,20;blue,38}, tail reversed, from=1-3, to=3-3]
	\arrow[tail reversed, from=2-1, to=1-2]
	\arrow[color={rgb,255:red,194;green,20;blue,38}, tail reversed, from=2-1, to=3-3]
	\arrow[color={rgb,255:red,194;green,20;blue,38}, tail reversed, from=3-2, to=2-4]
	\arrow[tail reversed, from=3-2, to=3-3]
\end{tikzcd}~,
\end{gathered}
\end{equation}
\begin{equation}
\begin{gathered}
v + \textcolor{VividBurgundy}{A}:
\begin{tikzcd}
	& 1 & v \\
	c &&& s \\
	& A & B
	\arrow[tail reversed, from=1-2, to=1-3]
	\arrow[color={rgb,255:red,194;green,20;blue,38}, tail reversed, from=1-2, to=3-2]
	\arrow[color={rgb,255:red,194;green,20;blue,38}, tail reversed, from=1-3, to=3-3]
	\arrow[tail reversed, from=2-1, to=2-4]
	\arrow[color={rgb,255:red,194;green,20;blue,38}, tail reversed, from=2-1, to=3-3]
	\arrow[color={rgb,255:red,194;green,20;blue,38}, tail reversed, from=3-2, to=2-4]
	\arrow[tail reversed, from=3-2, to=3-3]
\end{tikzcd}~.
\end{gathered}
\end{equation}
Here the coloring distinguishes the sub-quiver corresponding each subrepresentation. By inspection, the common sub-quiver of all minimal connected sub-quivers
\begin{equation}\tag{\ref{eq:Spin(9)_min_quiver}}
\begin{tikzcd}
	& v & 1 \\
	c & A & B & s
	\arrow[tail reversed, from=2-2, to=2-3]
\end{tikzcd},
\end{equation}
producing the required particle-soliton degeneracies discussed in the text.

\smallskip
\noindent $\bm{PSU(4)}+\psi_{\bm{15}}$. 
For the bosonized theory, the 3D coset TQFT is
\begin{equation}\tag{\ref{eq:PSU(4)_coset_TQFT}}
    \frac{Spin(15)_1 \times SU(4)_{-4}}{\Z_4} = \cZ(\cC).
\end{equation}
This theory has $14$ lines while its coset boundary condition has $4$, $\{1,v,A,B\}$. The condensation maps from the 3D theory to the coset boundary are:
\begin{equation}\label{eq:PSU(4)_projection_table}
\begin{tabular}{ |c|c| } 
\hline
$a$ & $p(a)$ \\
\hline
$(\mathbf{1}, \mathbf{1})$ &  $1$  \\
\hline
$(\mathbf{1}, \mathbf{35})$ &  $v$  \\
\hline
$(\mathbf{1}, \mathbf{45})$ &  $1 + A + B$ \\
\hline
$(\mathbf{1}, \mathbf{15})$ &  $v + A + B$ \\
\hline
$(\mathbf{1}, \mathbf{20}')_{1}$ & $A$ \\
\hline
$(\mathbf{1}, \mathbf{20}')_{2}$ &  $B$ \\
\hline
$(\mathbf{1}, \mathbf{84})_{1}$ & $B$ \\
\hline
\end{tabular} \quad 
\begin{tabular}{|c|c|}
\hline
$a$ & $p(a)$ \\
\hline
$(\mathbf{1}, \mathbf{84})_{2}$ & $A$ \\
\hline
$(\mathbf{s}, \mathbf{10})$ &  $A + B$ \\
\hline
$(\mathbf{s}, \mathbf{6})$ &  $A + B$ \\
\hline
$(\mathbf{s}, \mathbf{64})_{1}$ & $1 + B$ \\
\hline
$(\mathbf{s}, \mathbf{64})_{2}$ & $1 + A$ \\
\hline
$(\mathbf{s}, \mathbf{64})_{3}$ & $v + A$ \\
\hline
$(\mathbf{s}, \mathbf{64})_{4}$ & $v + B$ \\
\hline
\end{tabular}~.
\end{equation}
The boundary lines have non-trivial fusions
\begin{equation}\tag{\ref{eq:PSU(4)_fusion_table}}
\begin{tabular}{ |c|c|c|c| } 
\hline
$\times$  & $v$ & $A$ & $B$ \\
\hline
$v$ & $1$ & $B$ & $A$ \\ 
\hline
$A$ & $B$ & $1 + A + B$ & $v + A + B$ \\ 
\hline
$B$ & $A$ & $v + A + B$ &  $1 + A + B$ \\ 
\hline
\end{tabular}~.
\end{equation}
There are therefore $4$ distinct irreducible representations of the strip algebra. Following the prior two examples,  the module fusion coefficients are determined by \eqref{eq:PSU(4)_fusion_table}. 

As a final example, consider the line $A$. From \eqref{eq:PSU(4)_fusion_table}, the non-zero fusion coefficients we need are
\begin{equation}
\begin{gathered}
    N_{1,A}^A = N_{v,A}^B = 1, \\
    \begin{alignedat}{5}
        N_{A,A}^{1} &= N_{A,A}^A &&= N_{A,A}^B &&= 1, \\
        N_{B,A}^{v} &= N_{B,A}^A &&= N_{B,A}^B &&= 1.
    \end{alignedat}
\end{gathered}
\end{equation}
This defines the quiver
\begin{equation}\label{eq:PSU_4_A}
A: \quad
\begin{tikzcd}
	1 && B \\
	\\
	A && v
	\arrow[tail reversed, from=1-1, to=3-1]
	\arrow[from=1-3, to=1-3, loop, in=60, out=120, distance=5mm]
	\arrow[tail reversed, from=1-3, to=3-3]
	\arrow[tail reversed, from=3-1, to=1-3]
	\arrow[from=3-1, to=3-1, loop, in=240, out=300, distance=5mm]
\end{tikzcd}~.
\end{equation} 

Continuing in the same way, the remaining quivers are
\begin{equation}
1: \quad
\begin{tikzcd}
	1 && B \\
	\\
	A && v
	\arrow[from=1-1, to=1-1, loop, in=60, out=120, distance=5mm]
	\arrow[from=1-3, to=1-3, loop, in=60, out=120, distance=5mm]
	\arrow[from=3-1, to=3-1, loop, in=60, out=120, distance=5mm]
	\arrow[from=3-3, to=3-3, loop, in=60, out=120, distance=5mm]
\end{tikzcd}~,
\end{equation} 

\begin{equation}
v: \quad
\begin{tikzcd}
	1 && B \\
	\\
	A && v
	\arrow[tail reversed, from=1-1, to=3-3]
	\arrow[tail reversed, from=3-1, to=1-3]
\end{tikzcd}~,
\end{equation} 

\begin{equation}\label{eq:PSU_4_B}
B: \quad
\begin{tikzcd}
	1 && B \\
	\\
	A && v
	\arrow[tail reversed, from=1-1, to=1-3]
	\arrow[from=1-3, to=1-3, loop, in=60, out=120, distance=5mm]
	\arrow[tail reversed, from=3-1, to=1-3]
	\arrow[from=3-1, to=3-1, loop, in=240, out=300, distance=5mm]
	\arrow[tail reversed, from=3-1, to=3-3]
\end{tikzcd}~.
\end{equation} 
The minimal connected sub-quivers are easily seen to only be \eqref{eq:PSU_4_A} and \eqref{eq:PSU_4_B}. Their common connected sub-quiver is
\begin{equation}\tag{\ref{eq:PSU(4)_min_quiver}}
\begin{tikzcd}
	1 & A & B & v
	\arrow[from=1-2, to=1-2, loop, in=60, out=120, distance=5mm]
	\arrow[tail reversed, from=1-2, to=1-3]
	\arrow[from=1-3, to=1-3, loop, in=60, out=120, distance=5mm]
\end{tikzcd}~,
\end{equation}
implying the existence of degenerate particles and solitons as discussed in the main text.

\section{Vacuum Condensates} \label{VacuumCondensates}

In this section, we briefly discuss the characterization of the different vacua by the expectation values of the local operators that do not decouple in the IR. In the context of $\QCD$, the flow of local operators has been studied e.g. in \cite{Delmastro:2021otj, Delmastro:2022prj} (see also \cite{Cordova:2024goh}). As discussed in the main text, the topological local operators $\phi_{a}$ (in the basis diagonalized by the zero-form symmetry throughout the flow) can be obtained via anyon condensation by studying the endpoints of anyons at the topological boundaries. See \eqref{toplocaloperatorviaLagrangian}. In particular, we are interested in topological cosets where for simply-connected gauge group the Lagrangian algebra is determined by the branching rules
\begin{equation}
    \chi^{\mathcal{Q}_{1}}_{\Lambda}(q) = \sum_{\lambda} b_{(\Lambda, \lambda)} \, \chi^{G_{k}}_{\lambda}(q), \quad b_{(\Lambda, \lambda)}  \in \mathbb{N}.
\end{equation}
Specifically, $\mathcal{L} = \bigoplus_{(\Lambda, \lambda)} b_{(\Lambda, \lambda)} (\Lambda, \lambda^{\mathrm{op}})$. Thus, in the local operators $\phi_{a}$ the label $a$ runs through all the representations $(\Lambda, \lambda)$ such that $b_{(\Lambda, \lambda)} = 1$, since these would be the allowed endpoints in \eqref{toplocaloperatorviaLagrangian}. \footnote{Recall that in this work, for simplicity, we assume single multiplicity in the Lagrangian algebra. For higher multiplicity $b_{(\Lambda, \lambda)} \in \mathbb{Z}_{+}$, we must keep track of the multiple topological junctions at each boundary, and how these junctions can couple to each other in between the two boundaries of \eqref{toplocaloperatorviaLagrangian} in order to properly label local operators. See e.g.\ \cite{Cordova:2024goh} for a more detailed discussion on this issue.} As in Appendix \ref{app:example_calcs}, when the gauge group is not simply-connected, we denote the labels $a$ in the Lagrangian algebra of $(\mathcal{Q}_{1} \times G_{-k})/Z$ with $Z$ the ``common center'' in terms of their representatives in the simply-connected case before gauging $Z$.

We aim to diagnose the different vacua by the expectation value (condensates) of the operators $\phi_{a}$ in the presence of boundary conditions $i$. This can be done in terms of boundary states $| \mathbf{v}_{i} \rangle$:
\begin{equation}
    \langle \phi_{a} \rangle_{i} = \langle \mathbf{v}_{i} | \phi_{a} | \mathbf{v}_{i} \rangle.
\end{equation}
The overlap can be found using commutativity of the topological OPE of the $\phi_{a}$, associativity, consistency of the OPE with the junctions allowed by the Lagrangian algebra, and the well-known result that topological local operators always allow for an idempotent complete basis $\mathbf{v}_{i}$:
\begin{equation}
    \mathbf{v}_{i} \mathbf{v}_{j} = \delta_{ij} \mathbf{v}_{i}.
\end{equation}
The idempotent complete basis of local operators corresponds, actually, by the operator-state correspondence to the clustering boundary states $| \mathbf{v}_{i} \rangle$. The overlap is thus straightforwardly calculated using this result, and the expectation value of the order parameters can be encapsulated in a matrix
\begin{equation}
    \mathrm{B}_{a i} \coloneqq \langle \mathbf{v}_{i} | \phi_{a} | \mathbf{v}_{i} \rangle,
\end{equation}
providing the expectation value of the order parameter $\phi_{a}$ in the clustering vacuum state $|\mathbf{v}_{i} \rangle$.

\smallskip
\noindent $\bm{SO(3)}+\psi_{\bm5}$. Using the condensation data above for this example, we find that the idempotent complete basis is given as
\begin{align}
    \mathbf{v}_{1} &= \frac{1}{2(3 + \sqrt{3})} \big( \phi_{(\mathbf{1}, \mathbf{1})} + \phi_{(\mathbf{1}, \mathbf{7})} + \phi_{(\mathbf{s}, \mathbf{4})} \big), \label{SO31} \\[0.3cm]
    \mathbf{v}_{v} &= \frac{1}{2(3 + \sqrt{3})} \big( \phi_{(\mathbf{1}, \mathbf{1})} + \phi_{(\mathbf{1}, \mathbf{7})} - \phi_{(\mathbf{s}, \mathbf{4})} \big),  \\[0.3cm]
    \mathbf{v}_{A} &= \frac{1}{2\sqrt{3}} \Big( (1 + \sqrt{3}) \phi_{(\mathbf{1}, \mathbf{1})} + (1 - \sqrt{3}) \phi_{(\mathbf{1}, \mathbf{7})} \Big). \label{SO33}
\end{align}
Inverting \eqref{SO31}-\eqref{SO33}, we obtain the matrix of condensates:
\begin{equation}
\mathrm{B} = 
\begin{pmatrix}
1 & 1 & 1  \\
2 + \sqrt{3} & 2 + \sqrt{3} & -1 \\
3 + \sqrt{3} & - 3 - \sqrt{3} & 0  \\
\end{pmatrix}.
\end{equation}

\noindent $\bm{Spin(9)}+\psi_{\sigma}$. Using the condensation data above for this example, we find that the idempotent complete basis is given as
\begin{align}
    \mathbf{v}_{1} &= \frac{1}{12} \big( \phi_{(\mathbf{1},\mathbf{1})} + \phi_{(\mathbf{s},\mathbf{44})} + \phi_{(0,\mathbf{84})} \nonumber \\ & \hspace{2.0cm} + \phi_{(\mathbf{s}, \mathbf{84})} + \phi_{(\mathbf{v},\mathbf{16})} + \phi_{(\mathbf{c},\mathbf{128})} \big),  \label{Spin91} \\[0.3cm]
    \mathbf{v}_{s} &= \frac{1}{12} \big( \phi_{(\mathbf{1},\mathbf{1})} + \phi_{(\mathbf{s},\mathbf{44})} + \phi_{(0,\mathbf{84})} \nonumber \\ & \hspace{2.0cm} + \phi_{(\mathbf{s}, \mathbf{84})} - \phi_{(\mathbf{v},\mathbf{16})} - \phi_{(\mathbf{c},\mathbf{128})} \big),  \\[0.3cm]
    \mathbf{v}_{v} &= \frac{1}{12} \big( \phi_{(\mathbf{1},\mathbf{1})} - \phi_{(\mathbf{s},\mathbf{44})} + \phi_{(0,\mathbf{84})} \nonumber \\ & \hspace{2.0cm} - \phi_{(\mathbf{s}, \mathbf{84})} + \phi_{(\mathbf{v},\mathbf{16})} - \phi_{(\mathbf{c},\mathbf{128})} \big), \\[0.3cm]
    \mathbf{v}_{c} &= \frac{1}{12} \big( \phi_{(\mathbf{1},\mathbf{1})} - \phi_{(\mathbf{s},\mathbf{44})} + \phi_{(0,\mathbf{84})} \nonumber \\ & \hspace{2.0cm} - \phi_{(\mathbf{s}, \mathbf{84})} - \phi_{(\mathbf{v},\mathbf{16})}+ \phi_{(\mathbf{c},\mathbf{128})} \big), \\[0.3cm]
    \mathbf{v}_{A} &= \frac{1}{6} \big(  2 \phi_{(\mathbf{1},\mathbf{1})} + 2\phi_{(\mathbf{s},\mathbf{44})} - \phi_{(0,\mathbf{84})} - \phi_{(\mathbf{s}, \mathbf{84})} \big), \\[0.3cm]
    \mathbf{v}_{B} &= \frac{1}{6} \big( 2 \phi_{(\mathbf{1},\mathbf{1})} - 2\phi_{(\mathbf{s},\mathbf{44})} - \phi_{(0,\mathbf{84})} + \phi_{(\mathbf{s}, \mathbf{84})} \big). \label{Spin96}
\end{align}
Inverting \eqref{Spin91}-\eqref{Spin96}, we obtain the matrix of condensates:
\begin{equation}
\mathrm{B} = 
\begin{pmatrix}
1 & 1 & 1 & 1 & 1 & 1 \\
1 & 1 & -1 & -1 & 1 & -1 \\
2 & 2 & 2 & 2  & -1 & -1 \\
2 & 2 & -2 & -2  & -1 & 1 \\
3 & -3 & 3 & -3  & 0 & 0 \\
3 & -3 & -3 & 3  & 0 & 0 \\
\end{pmatrix}.
\end{equation}

\noindent $\bm{PSU(4)}+\psi_{\bm{15}}$. Using the condensation data above for this example, we find that the idempotent complete basis is given as
\begin{align}
    \mathbf{v}_{1} &= \frac{1}{4(2 + \sqrt{2})} \big( \phi_{(\mathbf{1}, \mathbf{1})} + \phi_{(\mathbf{1}, \mathbf{45})} \nonumber \\ & \hspace{3cm} + \phi_{(\mathbf{s}, \mathbf{64})_{1}} + \phi_{(\mathbf{s}, \mathbf{64})_{2}} \big), \label{PSU41} \\[0.3cm]
    \mathbf{v}_{v} &= \frac{1}{4(2 + \sqrt{2})} \big( \phi_{(\mathbf{1}, \mathbf{1})} + \phi_{(\mathbf{1}, \mathbf{45})} \nonumber \\ & \hspace{3cm} - \phi_{(\mathbf{s}, \mathbf{64})_{1}} - \phi_{(\mathbf{s}, \mathbf{64})_{2}} \big),  \\[0.3cm]
    \mathbf{v}_{A} &= \Big( (1 + \sqrt{2}) \phi_{(\mathbf{1}, \mathbf{1})} - (\sqrt{2} - 1) \phi_{(\mathbf{1}, \mathbf{45})} \nonumber \\ & \hspace{2cm} - \phi_{(\mathbf{s}, \mathbf{64})_{1}} + \phi_{(\mathbf{s}, \mathbf{64})_{2}}\Big)/ 4 \sqrt{2},  \\[0.3cm]
    \mathbf{v}_{B} &= \Big( (1 + \sqrt{2}) \phi_{(\mathbf{1}, \mathbf{1})} - (\sqrt{2} - 1) \phi_{(\mathbf{1}, \mathbf{45})} \nonumber \\ & \hspace{2cm} + \phi_{(\mathbf{s}, \mathbf{64})_{1}} - \phi_{(\mathbf{s}, \mathbf{64})_{2}}\Big)/ 4 \sqrt{2}. \label{PSU44}
\end{align}
Inverting \eqref{PSU41}-\eqref{PSU44}, we obtain the matrix of condensates:
\begin{equation}
\mathrm{B} = 
\begin{pmatrix}
1 & 1 & 1 & 1  \\
3 + 2 \sqrt{2} & 3 + 2 \sqrt{2} & -1 & -1  \\
2 + \sqrt{2} & - 2 - \sqrt{2} & -\sqrt{2} & \sqrt{2}   \\
2 + \sqrt{2} & - 2 - \sqrt{2} & \sqrt{2}  & -\sqrt{2} \\
\end{pmatrix}.
\end{equation}

\let\oldaddcontentsline\addcontentsline
\renewcommand{\addcontentsline}[3]{}
\bibliographystyle{apsrev4-1}
\bibliography{references}

\begin{thebibliography}{36}%
\makeatletter
\providecommand \@ifxundefined [1]{%
 \@ifx{#1\undefined}
}%
\providecommand \@ifnum [1]{%
 \ifnum #1\expandafter \@firstoftwo
 \else \expandafter \@secondoftwo
 \fi
}%
\providecommand \@ifx [1]{%
 \ifx #1\expandafter \@firstoftwo
 \else \expandafter \@secondoftwo
 \fi
}%
\providecommand \natexlab [1]{#1}%
\providecommand \enquote  [1]{``#1''}%
\providecommand \bibnamefont  [1]{#1}%
\providecommand \bibfnamefont [1]{#1}%
\providecommand \citenamefont [1]{#1}%
\providecommand \href@noop [0]{\@secondoftwo}%
\providecommand \href [0]{\begingroup \@sanitize@url \@href}%
\providecommand \@href[1]{\@@startlink{#1}\@@href}%
\providecommand \@@href[1]{\endgroup#1\@@endlink}%
\providecommand \@sanitize@url [0]{\catcode `\\12\catcode `\$12\catcode
  `\&12\catcode `\#12\catcode `\^12\catcode `\_12\catcode `\%12\relax}%
\providecommand \@@startlink[1]{}%
\providecommand \@@endlink[0]{}%
\providecommand \url  [0]{\begingroup\@sanitize@url \@url }%
\providecommand \@url [1]{\endgroup\@href {#1}{\urlprefix }}%
\providecommand \urlprefix  [0]{URL }%
\providecommand \Eprint [0]{\href }%
\providecommand \doibase [0]{http://dx.doi.org/}%
\providecommand \selectlanguage [0]{\@gobble}%
\providecommand \bibinfo  [0]{\@secondoftwo}%
\providecommand \bibfield  [0]{\@secondoftwo}%
\providecommand \translation [1]{[#1]}%
\providecommand \BibitemOpen [0]{}%
\providecommand \bibitemStop [0]{}%
\providecommand \bibitemNoStop [0]{.\EOS\space}%
\providecommand \EOS [0]{\spacefactor3000\relax}%
\providecommand \BibitemShut  [1]{\csname bibitem#1\endcsname}%
\let\auto@bib@innerbib\@empty
\bibitem [{\citenamefont {'t~Hooft}(1974)}]{tHooft:1974pnl}%
  \BibitemOpen
  \bibfield  {author} {\bibinfo {author} {\bibfnamefont {G.}~\bibnamefont
  {'t~Hooft}},\ }\href {\doibase 10.1016/0550-3213(74)90088-1} {\bibfield
  {journal} {\bibinfo  {journal} {Nucl. Phys. B}\ }\textbf {\bibinfo {volume}
  {75}},\ \bibinfo {pages} {461} (\bibinfo {year} {1974})}\BibitemShut
  {NoStop}%
\bibitem [{\citenamefont {Komargodski}\ \emph {et~al.}(2021)\citenamefont
  {Komargodski}, \citenamefont {Ohmori}, \citenamefont {Roumpedakis},\ and\
  \citenamefont {Seifnashri}}]{Komargodski:2020mxz}%
  \BibitemOpen
  \bibfield  {author} {\bibinfo {author} {\bibfnamefont {Z.}~\bibnamefont
  {Komargodski}}, \bibinfo {author} {\bibfnamefont {K.}~\bibnamefont {Ohmori}},
  \bibinfo {author} {\bibfnamefont {K.}~\bibnamefont {Roumpedakis}}, \ and\
  \bibinfo {author} {\bibfnamefont {S.}~\bibnamefont {Seifnashri}},\ }\href
  {\doibase 10.1007/JHEP03(2021)103} {\bibfield  {journal} {\bibinfo  {journal}
  {JHEP}\ }\textbf {\bibinfo {volume} {03}},\ \bibinfo {pages} {103} (\bibinfo
  {year} {2021})},\ \Eprint {http://arxiv.org/abs/2008.07567} {arXiv:2008.07567
  [hep-th]} \BibitemShut {NoStop}%
\bibitem [{\citenamefont {C\'{o}rdova}\ and\ \citenamefont
  {Garc\'\i{}a-Sep\'ulveda}(2024)}]{Cordova:2024goh}%
  \BibitemOpen
  \bibfield  {author} {\bibinfo {author} {\bibfnamefont {C.}~\bibnamefont
  {C\'{o}rdova}}\ and\ \bibinfo {author} {\bibfnamefont {D.}~\bibnamefont
  {Garc\'\i{}a-Sep\'ulveda}},\ }\href@noop {} {\  (\bibinfo {year} {2024})},\
  \Eprint {http://arxiv.org/abs/2412.01877} {arXiv:2412.01877 [hep-th]}
  \BibitemShut {NoStop}%
\bibitem [{\citenamefont {Delmastro}\ \emph {et~al.}(2023)\citenamefont
  {Delmastro}, \citenamefont {Gomis},\ and\ \citenamefont
  {Yu}}]{Delmastro:2021otj}%
  \BibitemOpen
  \bibfield  {author} {\bibinfo {author} {\bibfnamefont {D.}~\bibnamefont
  {Delmastro}}, \bibinfo {author} {\bibfnamefont {J.}~\bibnamefont {Gomis}}, \
  and\ \bibinfo {author} {\bibfnamefont {M.}~\bibnamefont {Yu}},\ }\href
  {\doibase 10.1007/JHEP02(2023)157} {\bibfield  {journal} {\bibinfo  {journal}
  {JHEP}\ }\textbf {\bibinfo {volume} {02}},\ \bibinfo {pages} {157} (\bibinfo
  {year} {2023})},\ \Eprint {http://arxiv.org/abs/2108.02202} {arXiv:2108.02202
  [hep-th]} \BibitemShut {NoStop}%
\bibitem [{\citenamefont {Delmastro}\ and\ \citenamefont
  {Gomis}(2023)}]{Delmastro:2022prj}%
  \BibitemOpen
  \bibfield  {author} {\bibinfo {author} {\bibfnamefont {D.}~\bibnamefont
  {Delmastro}}\ and\ \bibinfo {author} {\bibfnamefont {J.}~\bibnamefont
  {Gomis}},\ }\href {\doibase 10.1007/JHEP09(2023)158} {\bibfield  {journal}
  {\bibinfo  {journal} {JHEP}\ }\textbf {\bibinfo {volume} {09}},\ \bibinfo
  {pages} {158} (\bibinfo {year} {2023})},\ \Eprint
  {http://arxiv.org/abs/2211.09036} {arXiv:2211.09036 [hep-th]} \BibitemShut
  {NoStop}%
\bibitem [{\citenamefont {C\'{o}rdova}\ \emph
  {et~al.}(2024{\natexlab{a}})\citenamefont {C\'{o}rdova}, \citenamefont
  {Garc\'\i{}a-Sep\'ulveda},\ and\ \citenamefont
  {Holfester}}]{Cordova:2024vsq}%
  \BibitemOpen
  \bibfield  {author} {\bibinfo {author} {\bibfnamefont {C.}~\bibnamefont
  {C\'{o}rdova}}, \bibinfo {author} {\bibfnamefont {D.}~\bibnamefont
  {Garc\'\i{}a-Sep\'ulveda}}, \ and\ \bibinfo {author} {\bibfnamefont
  {N.}~\bibnamefont {Holfester}},\ }\href {\doibase 10.1007/JHEP07(2024)154}
  {\bibfield  {journal} {\bibinfo  {journal} {JHEP}\ }\textbf {\bibinfo
  {volume} {07}},\ \bibinfo {pages} {154} (\bibinfo {year}
  {2024}{\natexlab{a}})},\ \Eprint {http://arxiv.org/abs/2403.08883}
  {arXiv:2403.08883 [hep-th]} \BibitemShut {NoStop}%
\bibitem [{\citenamefont {C\'{o}rdova}\ \emph
  {et~al.}(2024{\natexlab{b}})\citenamefont {C\'{o}rdova}, \citenamefont
  {Holfester},\ and\ \citenamefont {Ohmori}}]{Cordova:2024iti}%
  \BibitemOpen
  \bibfield  {author} {\bibinfo {author} {\bibfnamefont {C.}~\bibnamefont
  {C\'{o}rdova}}, \bibinfo {author} {\bibfnamefont {N.}~\bibnamefont
  {Holfester}}, \ and\ \bibinfo {author} {\bibfnamefont {K.}~\bibnamefont
  {Ohmori}},\ }\href@noop {} {\  (\bibinfo {year} {2024}{\natexlab{b}})},\
  \Eprint {http://arxiv.org/abs/2408.11045} {arXiv:2408.11045 [hep-th]}
  \BibitemShut {NoStop}%
\bibitem [{\citenamefont {Frohlich}\ \emph {et~al.}(2007)\citenamefont
  {Frohlich}, \citenamefont {Fuchs}, \citenamefont {Runkel},\ and\
  \citenamefont {Schweigert}}]{Frohlich:2006ch}%
  \BibitemOpen
  \bibfield  {author} {\bibinfo {author} {\bibfnamefont {J.}~\bibnamefont
  {Frohlich}}, \bibinfo {author} {\bibfnamefont {J.}~\bibnamefont {Fuchs}},
  \bibinfo {author} {\bibfnamefont {I.}~\bibnamefont {Runkel}}, \ and\ \bibinfo
  {author} {\bibfnamefont {C.}~\bibnamefont {Schweigert}},\ }\href {\doibase
  10.1016/j.nuclphysb.2006.11.017} {\bibfield  {journal} {\bibinfo  {journal}
  {Nucl. Phys. B}\ }\textbf {\bibinfo {volume} {763}},\ \bibinfo {pages} {354}
  (\bibinfo {year} {2007})},\ \Eprint {http://arxiv.org/abs/hep-th/0607247}
  {arXiv:hep-th/0607247} \BibitemShut {NoStop}%
\bibitem [{\citenamefont {Frohlich}\ \emph {et~al.}(2010)\citenamefont
  {Frohlich}, \citenamefont {Fuchs}, \citenamefont {Runkel},\ and\
  \citenamefont {Schweigert}}]{Frohlich:2009gb}%
  \BibitemOpen
  \bibfield  {author} {\bibinfo {author} {\bibfnamefont {J.}~\bibnamefont
  {Frohlich}}, \bibinfo {author} {\bibfnamefont {J.}~\bibnamefont {Fuchs}},
  \bibinfo {author} {\bibfnamefont {I.}~\bibnamefont {Runkel}}, \ and\ \bibinfo
  {author} {\bibfnamefont {C.}~\bibnamefont {Schweigert}},\ }in\ \href
  {\doibase 10.1142/9789814304634_0056} {\emph {\bibinfo {booktitle} {{16th
  International Congress on Mathematical Physics}}}}\ (\bibinfo {year} {2010})\
  pp.\ \bibinfo {pages} {608--613},\ \Eprint {http://arxiv.org/abs/0909.5013}
  {arXiv:0909.5013 [math-ph]} \BibitemShut {NoStop}%
\bibitem [{\citenamefont {Bhardwaj}\ and\ \citenamefont
  {Tachikawa}(2018)}]{Bhardwaj:2017xup}%
  \BibitemOpen
  \bibfield  {author} {\bibinfo {author} {\bibfnamefont {L.}~\bibnamefont
  {Bhardwaj}}\ and\ \bibinfo {author} {\bibfnamefont {Y.}~\bibnamefont
  {Tachikawa}},\ }\href {\doibase 10.1007/JHEP03(2018)189} {\bibfield
  {journal} {\bibinfo  {journal} {JHEP}\ }\textbf {\bibinfo {volume} {03}},\
  \bibinfo {pages} {189} (\bibinfo {year} {2018})},\ \Eprint
  {http://arxiv.org/abs/1704.02330} {arXiv:1704.02330 [hep-th]} \BibitemShut
  {NoStop}%
\bibitem [{\citenamefont {Chang}\ \emph {et~al.}(2019)\citenamefont {Chang},
  \citenamefont {Lin}, \citenamefont {Shao}, \citenamefont {Wang},\ and\
  \citenamefont {Yin}}]{Chang:2018iay}%
  \BibitemOpen
  \bibfield  {author} {\bibinfo {author} {\bibfnamefont {C.-M.}\ \bibnamefont
  {Chang}}, \bibinfo {author} {\bibfnamefont {Y.-H.}\ \bibnamefont {Lin}},
  \bibinfo {author} {\bibfnamefont {S.-H.}\ \bibnamefont {Shao}}, \bibinfo
  {author} {\bibfnamefont {Y.}~\bibnamefont {Wang}}, \ and\ \bibinfo {author}
  {\bibfnamefont {X.}~\bibnamefont {Yin}},\ }\href {\doibase
  10.1007/JHEP01(2019)026} {\bibfield  {journal} {\bibinfo  {journal} {JHEP}\
  }\textbf {\bibinfo {volume} {01}},\ \bibinfo {pages} {026} (\bibinfo {year}
  {2019})},\ \Eprint {http://arxiv.org/abs/1802.04445} {arXiv:1802.04445
  [hep-th]} \BibitemShut {NoStop}%
\bibitem [{\citenamefont {Gaiotto}\ \emph {et~al.}(2015)\citenamefont
  {Gaiotto}, \citenamefont {Kapustin}, \citenamefont {Seiberg},\ and\
  \citenamefont {Willett}}]{Gaiotto:2014kfa}%
  \BibitemOpen
  \bibfield  {author} {\bibinfo {author} {\bibfnamefont {D.}~\bibnamefont
  {Gaiotto}}, \bibinfo {author} {\bibfnamefont {A.}~\bibnamefont {Kapustin}},
  \bibinfo {author} {\bibfnamefont {N.}~\bibnamefont {Seiberg}}, \ and\
  \bibinfo {author} {\bibfnamefont {B.}~\bibnamefont {Willett}},\ }\href
  {\doibase 10.1007/JHEP02(2015)172} {\bibfield  {journal} {\bibinfo  {journal}
  {JHEP}\ }\textbf {\bibinfo {volume} {02}},\ \bibinfo {pages} {172} (\bibinfo
  {year} {2015})},\ \Eprint {http://arxiv.org/abs/1412.5148} {arXiv:1412.5148
  [hep-th]} \BibitemShut {NoStop}%
\bibitem [{\citenamefont {Di~Francesco}\ and\ \citenamefont
  {Zuber}(1990)}]{DiFrancesco:1989ha}%
  \BibitemOpen
  \bibfield  {author} {\bibinfo {author} {\bibfnamefont {P.}~\bibnamefont
  {Di~Francesco}}\ and\ \bibinfo {author} {\bibfnamefont {J.~B.}\ \bibnamefont
  {Zuber}},\ }\href {\doibase 10.1016/0550-3213(90)90645-T} {\bibfield
  {journal} {\bibinfo  {journal} {Nucl. Phys. B}\ }\textbf {\bibinfo {volume}
  {338}},\ \bibinfo {pages} {602} (\bibinfo {year} {1990})}\BibitemShut
  {NoStop}%
\bibitem [{\citenamefont {Petkova}\ and\ \citenamefont
  {Zuber}(1996)}]{Petkova:1995fw}%
  \BibitemOpen
  \bibfield  {author} {\bibinfo {author} {\bibfnamefont {V.~B.}\ \bibnamefont
  {Petkova}}\ and\ \bibinfo {author} {\bibfnamefont {J.~B.}\ \bibnamefont
  {Zuber}},\ }\href {\doibase 10.1016/0550-3213(95)00670-2} {\bibfield
  {journal} {\bibinfo  {journal} {Nucl. Phys. B}\ }\textbf {\bibinfo {volume}
  {463}},\ \bibinfo {pages} {161} (\bibinfo {year} {1996})},\ \Eprint
  {http://arxiv.org/abs/hep-th/9510175} {arXiv:hep-th/9510175} \BibitemShut
  {NoStop}%
\bibitem [{\citenamefont {Behrend}\ \emph {et~al.}(2000)\citenamefont
  {Behrend}, \citenamefont {Pearce}, \citenamefont {Petkova},\ and\
  \citenamefont {Zuber}}]{Behrend:1999bn}%
  \BibitemOpen
  \bibfield  {author} {\bibinfo {author} {\bibfnamefont {R.~E.}\ \bibnamefont
  {Behrend}}, \bibinfo {author} {\bibfnamefont {P.~A.}\ \bibnamefont {Pearce}},
  \bibinfo {author} {\bibfnamefont {V.~B.}\ \bibnamefont {Petkova}}, \ and\
  \bibinfo {author} {\bibfnamefont {J.-B.}\ \bibnamefont {Zuber}},\ }\href
  {\doibase 10.1016/S0550-3213(99)00592-1} {\bibfield  {journal} {\bibinfo
  {journal} {Nucl. Phys. B}\ }\textbf {\bibinfo {volume} {570}},\ \bibinfo
  {pages} {525} (\bibinfo {year} {2000})},\ \Eprint
  {http://arxiv.org/abs/hep-th/9908036} {arXiv:hep-th/9908036} \BibitemShut
  {NoStop}%
\bibitem [{\citenamefont {Kirillov~Jr}\ and\ \citenamefont
  {Ostrik}(2002)}]{kirillov2002q}%
  \BibitemOpen
  \bibfield  {author} {\bibinfo {author} {\bibfnamefont {A.}~\bibnamefont
  {Kirillov~Jr}}\ and\ \bibinfo {author} {\bibfnamefont {V.}~\bibnamefont
  {Ostrik}},\ }\href@noop {} {\bibfield  {journal} {\bibinfo  {journal}
  {Advances in Mathematics}\ }\textbf {\bibinfo {volume} {171}},\ \bibinfo
  {pages} {183} (\bibinfo {year} {2002})}\BibitemShut {NoStop}%
\bibitem [{\citenamefont {Hung}\ and\ \citenamefont
  {Wan}(2015)}]{Hung:2015hfa}%
  \BibitemOpen
  \bibfield  {author} {\bibinfo {author} {\bibfnamefont {L.-Y.}\ \bibnamefont
  {Hung}}\ and\ \bibinfo {author} {\bibfnamefont {Y.}~\bibnamefont {Wan}},\
  }\href {\doibase 10.1007/JHEP07(2015)120} {\bibfield  {journal} {\bibinfo
  {journal} {JHEP}\ }\textbf {\bibinfo {volume} {07}},\ \bibinfo {pages} {120}
  (\bibinfo {year} {2015})},\ \Eprint {http://arxiv.org/abs/1502.02026}
  {arXiv:1502.02026 [cond-mat.str-el]} \BibitemShut {NoStop}%
\bibitem [{\citenamefont {Cecotti}\ and\ \citenamefont
  {Vafa}(1993)}]{Cecotti:1992rm}%
  \BibitemOpen
  \bibfield  {author} {\bibinfo {author} {\bibfnamefont {S.}~\bibnamefont
  {Cecotti}}\ and\ \bibinfo {author} {\bibfnamefont {C.}~\bibnamefont {Vafa}},\
  }\href {\doibase 10.1007/BF02096804} {\bibfield  {journal} {\bibinfo
  {journal} {Commun. Math. Phys.}\ }\textbf {\bibinfo {volume} {158}},\
  \bibinfo {pages} {569} (\bibinfo {year} {1993})},\ \Eprint
  {http://arxiv.org/abs/hep-th/9211097} {arXiv:hep-th/9211097} \BibitemShut
  {NoStop}%
\bibitem [{\citenamefont {Witten}(1984)}]{Witten:1983ar}%
  \BibitemOpen
  \bibfield  {author} {\bibinfo {author} {\bibfnamefont {E.}~\bibnamefont
  {Witten}},\ }\href {\doibase 10.1007/BF01215276} {\bibfield  {journal}
  {\bibinfo  {journal} {Commun. Math. Phys.}\ }\textbf {\bibinfo {volume}
  {92}},\ \bibinfo {pages} {455} (\bibinfo {year} {1984})}\BibitemShut
  {NoStop}%
\bibitem [{\citenamefont {Ji}\ \emph {et~al.}(2020)\citenamefont {Ji},
  \citenamefont {Shao},\ and\ \citenamefont {Wen}}]{Ji:2019ugf}%
  \BibitemOpen
  \bibfield  {author} {\bibinfo {author} {\bibfnamefont {W.}~\bibnamefont
  {Ji}}, \bibinfo {author} {\bibfnamefont {S.-H.}\ \bibnamefont {Shao}}, \ and\
  \bibinfo {author} {\bibfnamefont {X.-G.}\ \bibnamefont {Wen}},\ }\href
  {\doibase 10.1103/PhysRevResearch.2.033317} {\bibfield  {journal} {\bibinfo
  {journal} {Phys. Rev. Res.}\ }\textbf {\bibinfo {volume} {2}},\ \bibinfo
  {pages} {033317} (\bibinfo {year} {2020})},\ \Eprint
  {http://arxiv.org/abs/1909.01425} {arXiv:1909.01425 [cond-mat.str-el]}
  \BibitemShut {NoStop}%
\bibitem [{\citenamefont {Thorngren}\ and\ \citenamefont
  {Wang}(2024{\natexlab{a}})}]{Thorngren:2021yso}%
  \BibitemOpen
  \bibfield  {author} {\bibinfo {author} {\bibfnamefont {R.}~\bibnamefont
  {Thorngren}}\ and\ \bibinfo {author} {\bibfnamefont {Y.}~\bibnamefont
  {Wang}},\ }\href {\doibase 10.1007/JHEP07(2024)051} {\bibfield  {journal}
  {\bibinfo  {journal} {JHEP}\ }\textbf {\bibinfo {volume} {07}},\ \bibinfo
  {pages} {051} (\bibinfo {year} {2024}{\natexlab{a}})},\ \Eprint
  {http://arxiv.org/abs/2106.12577} {arXiv:2106.12577 [hep-th]} \BibitemShut
  {NoStop}%
\bibitem [{\citenamefont {C\'{o}rdova}\ and\ \citenamefont
  {Garc\'\i{}a-Sep\'ulveda}(2023)}]{Cordova:2023jip}%
  \BibitemOpen
  \bibfield  {author} {\bibinfo {author} {\bibfnamefont {C.}~\bibnamefont
  {C\'{o}rdova}}\ and\ \bibinfo {author} {\bibfnamefont {D.}~\bibnamefont
  {Garc\'\i{}a-Sep\'ulveda}},\ }\href@noop {} {\  (\bibinfo {year} {2023})},\
  \Eprint {http://arxiv.org/abs/2312.16317} {arXiv:2312.16317 [hep-th]}
  \BibitemShut {NoStop}%
\bibitem [{\citenamefont {Damia}\ \emph {et~al.}(2024)\citenamefont {Damia},
  \citenamefont {Galati},\ and\ \citenamefont {Tizzano}}]{Damia:2024kyt}%
  \BibitemOpen
  \bibfield  {author} {\bibinfo {author} {\bibfnamefont {J.~A.}\ \bibnamefont
  {Damia}}, \bibinfo {author} {\bibfnamefont {G.}~\bibnamefont {Galati}}, \
  and\ \bibinfo {author} {\bibfnamefont {L.}~\bibnamefont {Tizzano}},\
  }\href@noop {} {\  (\bibinfo {year} {2024})},\ \Eprint
  {http://arxiv.org/abs/2409.17989} {arXiv:2409.17989 [hep-th]} \BibitemShut
  {NoStop}%
\bibitem [{\citenamefont {Kong}(2014)}]{Kong:2013aya}%
  \BibitemOpen
  \bibfield  {author} {\bibinfo {author} {\bibfnamefont {L.}~\bibnamefont
  {Kong}},\ }\href {\doibase 10.1016/j.nuclphysb.2014.07.003} {\bibfield
  {journal} {\bibinfo  {journal} {Nucl. Phys. B}\ }\textbf {\bibinfo {volume}
  {886}},\ \bibinfo {pages} {436} (\bibinfo {year} {2014})},\ \Eprint
  {http://arxiv.org/abs/1307.8244} {arXiv:1307.8244 [cond-mat.str-el]}
  \BibitemShut {NoStop}%
\bibitem [{\citenamefont {Kutasov}\ and\ \citenamefont
  {Schwimmer}(1995)}]{Kutasov:1994xq}%
  \BibitemOpen
  \bibfield  {author} {\bibinfo {author} {\bibfnamefont {D.}~\bibnamefont
  {Kutasov}}\ and\ \bibinfo {author} {\bibfnamefont {A.}~\bibnamefont
  {Schwimmer}},\ }\href {\doibase 10.1016/0550-3213(95)00106-3} {\bibfield
  {journal} {\bibinfo  {journal} {Nucl. Phys. B}\ }\textbf {\bibinfo {volume}
  {442}},\ \bibinfo {pages} {447} (\bibinfo {year} {1995})},\ \Eprint
  {http://arxiv.org/abs/hep-th/9501024} {arXiv:hep-th/9501024} \BibitemShut
  {NoStop}%
\bibitem [{\citenamefont {Kitaev}(2006)}]{Kitaev:2005hzj}%
  \BibitemOpen
  \bibfield  {author} {\bibinfo {author} {\bibfnamefont {A.}~\bibnamefont
  {Kitaev}},\ }\href {\doibase 10.1016/j.aop.2005.10.005} {\bibfield  {journal}
  {\bibinfo  {journal} {Annals Phys.}\ }\textbf {\bibinfo {volume} {321}},\
  \bibinfo {pages} {2} (\bibinfo {year} {2006})},\ \Eprint
  {http://arxiv.org/abs/cond-mat/0506438} {arXiv:cond-mat/0506438} \BibitemShut
  {NoStop}%
\bibitem [{\citenamefont {Benini}\ \emph {et~al.}(2019)\citenamefont {Benini},
  \citenamefont {C\'{o}rdova},\ and\ \citenamefont {Hsin}}]{Benini:2018reh}%
  \BibitemOpen
  \bibfield  {author} {\bibinfo {author} {\bibfnamefont {F.}~\bibnamefont
  {Benini}}, \bibinfo {author} {\bibfnamefont {C.}~\bibnamefont {C\'{o}rdova}},
  \ and\ \bibinfo {author} {\bibfnamefont {P.-S.}\ \bibnamefont {Hsin}},\
  }\href {\doibase 10.1007/JHEP03(2019)118} {\bibfield  {journal} {\bibinfo
  {journal} {JHEP}\ }\textbf {\bibinfo {volume} {03}},\ \bibinfo {pages} {118}
  (\bibinfo {year} {2019})},\ \Eprint {http://arxiv.org/abs/1803.09336}
  {arXiv:1803.09336 [hep-th]} \BibitemShut {NoStop}%
\bibitem [{\citenamefont {Kaidi}\ \emph {et~al.}(2022)\citenamefont {Kaidi},
  \citenamefont {Komargodski}, \citenamefont {Ohmori}, \citenamefont
  {Seifnashri},\ and\ \citenamefont {Shao}}]{Kaidi:2021gbs}%
  \BibitemOpen
  \bibfield  {author} {\bibinfo {author} {\bibfnamefont {J.}~\bibnamefont
  {Kaidi}}, \bibinfo {author} {\bibfnamefont {Z.}~\bibnamefont {Komargodski}},
  \bibinfo {author} {\bibfnamefont {K.}~\bibnamefont {Ohmori}}, \bibinfo
  {author} {\bibfnamefont {S.}~\bibnamefont {Seifnashri}}, \ and\ \bibinfo
  {author} {\bibfnamefont {S.-H.}\ \bibnamefont {Shao}},\ }\href {\doibase
  10.21468/SciPostPhys.13.3.067} {\bibfield  {journal} {\bibinfo  {journal}
  {SciPost Phys.}\ }\textbf {\bibinfo {volume} {13}},\ \bibinfo {pages} {067}
  (\bibinfo {year} {2022})},\ \Eprint {http://arxiv.org/abs/2107.13091}
  {arXiv:2107.13091 [hep-th]} \BibitemShut {NoStop}%
\bibitem [{\citenamefont {Davydov}\ \emph {et~al.}(2013)\citenamefont
  {Davydov}, \citenamefont {M{\"u}ger}, \citenamefont {Nikshych},\ and\
  \citenamefont {Ostrik}}]{davydov2013witt}%
  \BibitemOpen
  \bibfield  {author} {\bibinfo {author} {\bibfnamefont {A.}~\bibnamefont
  {Davydov}}, \bibinfo {author} {\bibfnamefont {M.}~\bibnamefont {M{\"u}ger}},
  \bibinfo {author} {\bibfnamefont {D.}~\bibnamefont {Nikshych}}, \ and\
  \bibinfo {author} {\bibfnamefont {V.}~\bibnamefont {Ostrik}},\ }\href@noop {}
  {\bibfield  {journal} {\bibinfo  {journal} {Journal f{\"u}r die reine und
  angewandte Mathematik (Crelles Journal)}\ }\textbf {\bibinfo {volume}
  {2013}},\ \bibinfo {pages} {135} (\bibinfo {year} {2013})}\BibitemShut
  {NoStop}%
\bibitem [{\citenamefont {Moore}\ and\ \citenamefont
  {Seiberg}(1989)}]{Moore:1989yh}%
  \BibitemOpen
  \bibfield  {author} {\bibinfo {author} {\bibfnamefont {G.~W.}\ \bibnamefont
  {Moore}}\ and\ \bibinfo {author} {\bibfnamefont {N.}~\bibnamefont
  {Seiberg}},\ }\href {\doibase 10.1016/0370-2693(89)90897-6} {\bibfield
  {journal} {\bibinfo  {journal} {Phys. Lett. B}\ }\textbf {\bibinfo {volume}
  {220}},\ \bibinfo {pages} {422} (\bibinfo {year} {1989})}\BibitemShut
  {NoStop}%
\bibitem [{\citenamefont {Hori}(1996)}]{Hori:1994nc}%
  \BibitemOpen
  \bibfield  {author} {\bibinfo {author} {\bibfnamefont {K.}~\bibnamefont
  {Hori}},\ }\href {\doibase 10.1007/BF02506384} {\bibfield  {journal}
  {\bibinfo  {journal} {Commun. Math. Phys.}\ }\textbf {\bibinfo {volume}
  {182}},\ \bibinfo {pages} {1} (\bibinfo {year} {1996})},\ \Eprint
  {http://arxiv.org/abs/hep-th/9411134} {arXiv:hep-th/9411134} \BibitemShut
  {NoStop}%
\bibitem [{\citenamefont {Carqueville}\ and\ \citenamefont
  {Runkel}(2016)}]{Carqueville:2012Dk}%
  \BibitemOpen
  \bibfield  {author} {\bibinfo {author} {\bibfnamefont {N.}~\bibnamefont
  {Carqueville}}\ and\ \bibinfo {author} {\bibfnamefont {I.}~\bibnamefont
  {Runkel}},\ }\href {\doibase 10.4171/qt/76} {\bibfield  {journal} {\bibinfo
  {journal} {Quantum Topol.}\ }\textbf {\bibinfo {volume} {7}},\ \bibinfo
  {pages} {203} (\bibinfo {year} {2016})},\ \Eprint
  {http://arxiv.org/abs/1210.6363} {arXiv:1210.6363 [math.QA]} \BibitemShut
  {NoStop}%
\bibitem [{\citenamefont {Thorngren}\ and\ \citenamefont
  {Wang}(2024{\natexlab{b}})}]{Thorngren:2019iar}%
  \BibitemOpen
  \bibfield  {author} {\bibinfo {author} {\bibfnamefont {R.}~\bibnamefont
  {Thorngren}}\ and\ \bibinfo {author} {\bibfnamefont {Y.}~\bibnamefont
  {Wang}},\ }\href {\doibase 10.1007/JHEP04(2024)132} {\bibfield  {journal}
  {\bibinfo  {journal} {JHEP}\ }\textbf {\bibinfo {volume} {04}},\ \bibinfo
  {pages} {132} (\bibinfo {year} {2024}{\natexlab{b}})},\ \Eprint
  {http://arxiv.org/abs/1912.02817} {arXiv:1912.02817 [hep-th]} \BibitemShut
  {NoStop}%
\bibitem [{\citenamefont {Huang}\ \emph {et~al.}(2021)\citenamefont {Huang},
  \citenamefont {Lin},\ and\ \citenamefont {Seifnashri}}]{Huang:2021zvu}%
  \BibitemOpen
  \bibfield  {author} {\bibinfo {author} {\bibfnamefont {T.-C.}\ \bibnamefont
  {Huang}}, \bibinfo {author} {\bibfnamefont {Y.-H.}\ \bibnamefont {Lin}}, \
  and\ \bibinfo {author} {\bibfnamefont {S.}~\bibnamefont {Seifnashri}},\
  }\href {\doibase 10.1007/JHEP12(2021)028} {\bibfield  {journal} {\bibinfo
  {journal} {JHEP}\ }\textbf {\bibinfo {volume} {12}},\ \bibinfo {pages} {028}
  (\bibinfo {year} {2021})},\ \Eprint {http://arxiv.org/abs/2110.02958}
  {arXiv:2110.02958 [hep-th]} \BibitemShut {NoStop}%
\bibitem [{\citenamefont {Diatlyk}\ \emph {et~al.}(2024)\citenamefont
  {Diatlyk}, \citenamefont {Luo}, \citenamefont {Wang},\ and\ \citenamefont
  {Weller}}]{Diatlyk:2023fwf}%
  \BibitemOpen
  \bibfield  {author} {\bibinfo {author} {\bibfnamefont {O.}~\bibnamefont
  {Diatlyk}}, \bibinfo {author} {\bibfnamefont {C.}~\bibnamefont {Luo}},
  \bibinfo {author} {\bibfnamefont {Y.}~\bibnamefont {Wang}}, \ and\ \bibinfo
  {author} {\bibfnamefont {Q.}~\bibnamefont {Weller}},\ }\href {\doibase
  10.1007/JHEP03(2024)127} {\bibfield  {journal} {\bibinfo  {journal} {JHEP}\
  }\textbf {\bibinfo {volume} {03}},\ \bibinfo {pages} {127} (\bibinfo {year}
  {2024})},\ \Eprint {http://arxiv.org/abs/2311.17044} {arXiv:2311.17044
  [hep-th]} \BibitemShut {NoStop}%
\bibitem [{\citenamefont {Etingof}\ \emph {et~al.}(2015)\citenamefont
  {Etingof}, \citenamefont {Gelaki}, \citenamefont {Nikshych},\ and\
  \citenamefont {Ostrik}}]{ReferenceKey}%
  \BibitemOpen
  \bibfield  {author} {\bibinfo {author} {\bibfnamefont {P.}~\bibnamefont
  {Etingof}}, \bibinfo {author} {\bibfnamefont {S.}~\bibnamefont {Gelaki}},
  \bibinfo {author} {\bibfnamefont {D.}~\bibnamefont {Nikshych}}, \ and\
  \bibinfo {author} {\bibfnamefont {V.}~\bibnamefont {Ostrik}},\ }\href@noop {}
  {\emph {\bibinfo {title} {Tensor Categories: Mathematical Surveys and
  Monographs. American Mathematical Society}}}\ (\bibinfo  {publisher}
  {American Mathematical Society},\ \bibinfo {year} {2015})\BibitemShut
  {NoStop}%
\end{thebibliography}%
\let\addcontentsline\oldaddcontentsline

\end{document}